\documentclass[%
aps, %
prb, 
preprint, onecolumn, 11pt, 
a4paper, %
amsfonts, amsmath, amssymb]{revtex4-1}
\usepackage[sort&compress]{natbib}
\usepackage{graphicx}
\usepackage{mathrsfs}
\usepackage{bm}
\usepackage{braket}
\usepackage[dvipsnames]{xcolor}
\usepackage{hyperref}
\hypersetup{%
	colorlinks, %
	linkcolor=red, %
	citecolor=blue, %
	urlcolor=ForestGreen %
}
\newcommand{\br}{\bm{r}}
\newcommand{\bk}{\bm{k}}
\newcommand{\bp}{\bm{p}}
\newcommand{\bq}{\bm{q}}
\newcommand{\zi}{i}
\renewcommand{\braket}[1]{\left\langle #1 \right\rangle}

\newcommand{\kB}{k_{\mathrm{B}}}
\newcommand{\me}{m_{\rm e}}
\newcommand{\re}{\mathrm{Re}}
\newcommand{\im}{\mathrm{Im}}
\newcommand{\eF}{\epsilon_{\rm F}}
\allowdisplaybreaks[1]

\begin{document}
%
\title{Non-local Spin-charge Conversion via Rashba Spin-Orbit Interaction}
\date{\today}
\author{Junji Fujimoto}
\email[E-mail address: ]{fujimoto.junji.s8@kyoto-u.ac.jp}
\affiliation{Institute for Chemical Research, Kyoto University, Uji, Kyoto 611-0011, Japan}
\affiliation{RIKEN Center for Emergent Matter Science (CEMS), Wako, Saitama 351-0198, Japan}
\author{Gen Tatara}
\affiliation{RIKEN Center for Emergent Matter Science (CEMS) and RIKEN Cluster for Pioneering Research (CPR), Wako, Saitama 351-0198, Japan}
\begin{abstract}
We show theoretically that conversion between spin and charge by spin-orbit interaction in metals  occurs even in a non-local setup where magnetization and spin-orbit interaction are spatially separated if electron diffusion is taken into account.
Calculation is carried out for the Rashba spin-orbit interaction treating the coupling with a ferromagnet perturbatively.
The results indicate the validity of the concept of effective spin gauge field (spin motive force) in the non-local configuration.
The inverse Rashba-Edelstein effect observed for a trilayer of a ferromagnet, a normal metal and a heavy metal can be explained in terms of the non-local effective spin gauge field.
\end{abstract}
\maketitle

\section{\label{sec:introduction}Introduction}
The objective of spintronics is to manipulate spins by electric means and vice versa. 
For generating spin accumulation and spin current, several methods have been  experimentally established in the last two decades, including  the spin pumping effect~\cite{Silsbee1979,Mizukami2001,Tserkovnyak2002,Costache2006}, where magnetization precession of a ferromagnet (F) is used to generate spin current into a normal metal (NM) in a F-NM junction. 
For electric detection of spin current, so called the inverse spin Hall effect induced by spin-orbit interaction of heavy metal is widely used~\cite{Saitoh2006}.

\begin{figure}[t]
\includegraphics[width=0.4\textwidth]{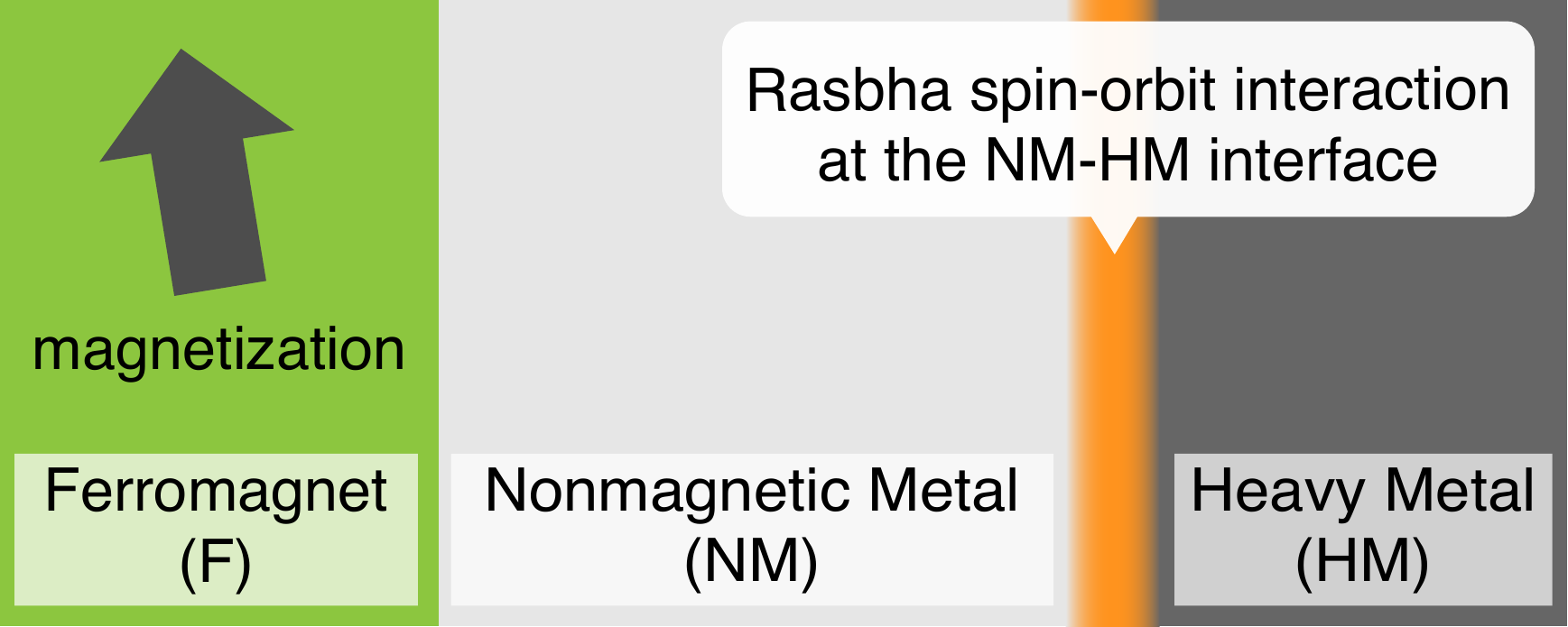}
\caption{Schematic figure of a trilayer of a ferromagnet (F), a normal metal (NM) and a heavy metal (HM).  The Rashba spin-orbit interaction is localized at the NM-HM interface.
 \label{figlayer}}
\end{figure}

Another electric detection of spin is by interfacial Rashba spin-orbit interaction, called the inverse Rashba-Edelstein effect.
The effect, the reciprocal effect of the current-induced spin polarization studied theoretically by Edelstein \cite{Edelstein90}, has been experimentally demonstrated in a trilayer of a ferromagnet, a normal metal and a heavy metal (HM) (Fig. \ref{figlayer}) \cite{Sanchez2013}.
The Rashba interaction is expected to be localized at the interface of NM and HM, and the normal metal works as a spacer to separate the magnetization and the Rashba interaction.
The current observed was argued to support the spin current picture, in which a spin current generated by spin pumping effect propagates through the normal metal, forming spin accumulation at the NM-HM interface, finally resulting in a current as a result of  inverse Rashba-Edelstein effect.

Theoretically, current generation indicates existence of an effective electric field (motive force).
In the present magnetic systems, it is the one driving electron spin, namely, effective spin gauge field or spin motive force.
Effective spin gauge field has been known to arise for a slowly-varying magnetic texture in ferromagnets, as the texture gives rise to a phase (spin Berry's phase) for electron spin wave function as a result of $sd$ exchange interaction \cite{Volovik1987,TataraReview19}.
The spin Berry's phase generates a current when magnetization has dynamics besides spatial texture. 
The concept of effective spin gauge field was shown to be generalized to include spin-orbit interactions that is linear in the wave vector, like the Rashba interaction  \cite{Kim2012,Takeuchi2012,Nakabayashi2014}.
It was demonstrated that magnetization $\bm{M}$ and the Rashba field $\bm{\alpha}$ give rise to an effective spin gauge field proportional to $\bm{\alpha}\times \bm{M}$, leading to an effective electric field and current proportional to $\bm{\alpha}\times \dot{\bm{M}}$. It was also pointed out that spin relaxation leads to a perpendicular effective spin electric field,  
$\bm{\alpha}\times(\bm{M}\times \dot{\bm{M}})$ \cite{Tatara_smf13}. 
The latter component leads to a direct current~(DC) for a precessing magnetization, while the former induces only alternating current~(AC).
The DC is in the direction perpendicular to both $\bm{\alpha}$ and the average of $\bm{M}$, which agrees with the geometry of the experimentally-observed inverse Rashba-Edelstein effect. 
The above theories, however, do not directly apply to the experimental situations with a spacer layer, as the coexistence of magnetization and Rashba interaction is assumed in the theories.

The objective of the paper is to demonstrate theoretically that the concept of effective electric field can be generalized to describe non-local configurations where the magnetization and the Rashba interaction are spatially separated by a nonmagnetic metal. 
For charge transport in metals, longer distance than the electron mean free path is possible due to electron diffusion; Current generation in disordered metals can therefore occur non-locally.
As far as the diffusion is induced by elastic scattering conserving spin, the spin information is expected to be equally transported long distance.
In fact, long-range diffusive component of spin current induced by magnetization dynamics was studied in Ref. \cite{Takeuchi10}.
The spin-charge conversion effect was briefly mentioned there, but assuming uniform Rashba interaction.
It was also pointed out that long-range spin chirality contributes to the anomalous Hall effect in disordered ferromagnet if the spatial size is less than spin diffusion length \cite{Nakazawa2014}, indicating that spin Berry's phase has long-range components. 
Moreover, spin Hall and inverse spin Hall effects were recently formulated in terms of non-local conversion of spin and electric current including electron diffusion \cite{Tatara18}.
In this representation, the observed spin density (or current) in the direct (inverse) spin Hall effect is directly related to the driving electric field (or spin pumping field) via a non-local response function of spin and electric current.

In this paper, we calculate electric current generated in the system of conduction electron with $sd$ exchange interaction with a dynamic magnetization and the Rashba interaction having spatial distributions. 
Although experiments are carried out in trilayers with rather sharp interfaces, we here describe the slowly-varying case, which can be accessible straightforwardly by an expansion with respect to the wave vectors of the two interactions. 
The vertex corrections~(VCs) to the correlation function representing the effect turns out to contain a singular pole at slowly-varying limit, indicating the diffusive nature. 
The diffusion propagator arising from this pole is shown to connect the information of the magnetization and the Rashba interaction even when they are spatially separated, resulting in a non-local current generation.
The effect is interpreted in terms of the non-local component of spin electric field.

\section{\label{sec:model_Green-functions}Model and Green functions}
 We consider the following model Hamiltonian, $\mathscr{H} = \mathcal{H}_{0} + \mathcal{H}_{\rm R} + \mathcal{H}_{sd} (t)$ with
\begin{subequations}
\begin{align}
\mathcal{H}_{0}
	& = \int \mathrm{d}\br\, \psi^{\dagger} (\br) \left(
		- \frac{\hbar^2 \bm{\nabla}^2}{2 \me}
		+ u V \sum_{i = 1}^{N_{\rm i}} \delta ( \br - \bm{R}_i )
		\right) \psi (\br)
, \\
\mathcal{H}_{\rm R}
	& = - \frac{\zi \hbar}{2} \epsilon^{i j k} \int \mathrm{d}\br\, \alpha_{i} (\br) \left(
			\psi^{\dagger} (\br) \sigma_k \bigl( \nabla_{j} \psi (\br) \bigr)
			- \bigl( \nabla_{j} \psi^{\dagger} (\br) \bigr) \sigma_k \psi (\br)
		\right)
, \\
\mathcal{H}_{sd} (t)
	& = - \int \mathrm{d}\br\, \psi^{\dagger} (\br) \bigl(
		\bm{M} (\br, t) \cdot \bm{\sigma}
		\bigr) \psi (\br)
,\end{align}
\label{eq:Hamiltonian_r}%
\end{subequations}
where $\psi^{} (\br) = {}^{\rm t} (\psi_{\uparrow} (\br), \, \psi_{\downarrow} (\br))$ is the spinor form of the annihilation operator of electron with the mass being $\me$, and $\mathcal{H}_{0}$ consists of the kinetic term and non-magnetic impurity potential with the strength being $u$.
 $\mathcal{H}_{\rm R}$ is the Rashba spin-orbit interaction with the spatial-dependent Rashba field denoted by $\bm{\alpha} (\br) = (\alpha_x (\br), \alpha_y (\br), \alpha_z (\br))$.
 The $sd$ exchange interaction is given by $\mathcal{H}_{sd} (t)$, where $\bm{M} (\br, t)$ is the magnetization vector including the $sd$ interaction strength.
 We deal with $\mathcal{H}_0$ as the unperturbed Hamiltonian and treat $\mathcal{H}_{\rm R}$ and $\mathcal{H}_{sd}$ perturbatively.
 Here, $V$ is the volume of the system, $\bm{\sigma} = (\sigma_x, \sigma_y, \sigma_z)$ is the vector form of the Pauli matrices, $\hbar$ is the Planck constant divided by $2 \pi$, and $\epsilon^{i j k}$ is the Levi-Civita symbol.
We consider slowly-varying case with weak spatial dependencies of $M (\br)$ and $\bm{\alpha} (\br)$.  
 Our particular interest is the case where $M (\br)$ and $\bm{\alpha} (\br)$ do not coexist, such as a tri-layer structure composed of F, NM and HM.
 
 The charge current density operator of the system is given by
\begin{align}
\bm{j} (\br)
	& = - \frac{e \hbar}{2 \me \zi}
		\left(
			\psi^{\dagger} (\br) \bigl( \bm{\nabla} \psi^{} (\br) \bigr)
			- \bigl( \bm{\nabla} \psi^{\dagger} (\br) \bigr) \psi^{} (\br)
		\right)
		+ e \bm{\alpha} (\br) \times \bm{s} (\br)
\label{eq:current_density_def}
,\end{align}
the first two terms of which we call the current density for the normal velocity and the last term is called that for the anomalous velocity, where $e \,(>0)$ is the elementary charge, and $\bm{s} (\br) = \psi^{\dagger} (\br) \bm{\sigma} \psi^{} (\br)$ is the spin density.
 In the Fourier forms, the Hamiltonians of Eqs.~(\ref{eq:Hamiltonian_r}) are given as
\begin{subequations}
\begin{align}
\mathcal{H}_0
	& = \sum_{\bk} \epsilon_{\bk} c^{\dagger}_{\bk} c^{}_{\bk}
		+ u \sum_{\bk, \bq} \sum_{i = 1}^{N_{\rm i}} e^{\zi \bq \cdot \bm{R}_i}
			c^{\dagger}_{\bk+\frac{\bq}{2}} c^{}_{\bk-\frac{\bq}{2}}
, \\
\mathcal{H}_{sd} (t)
	& = - \int \frac{\mathrm{d}\omega}{2 \pi} e^{-\zi \omega t} \sum_{\bq}
		\bm{S}(-\bq) \cdot \bm{M} (\bq, \omega)
, \\
\mathcal{H}_{\rm R}
	& = \sum_{\bk, \bq}
		(\bm{\alpha}_{\bq} \times \hbar \bk) \cdot c^{\dagger}_{\bk+\frac{\bq}{2}} \bm{\sigma} c^{}_{\bk-\frac{\bq}{2}}
,\end{align}
\end{subequations}
where $\epsilon_{\bk} = \hbar^2 k^2 / 2 \me$, and $\bm{s}(\bq)$ is the Fourier component of the spin operator given by
\begin{align}
\bm{s} (\bq)
	& = \frac{1}{V} \sum_{\bk} c^{\dagger}_{\bk-\frac{\bq}{2}}  \bm{\sigma} c^{}_{\bk+\frac{\bq}{2}}
\end{align}
 The current density in the Fourier form is given as
\begin{align}
\bm{j} (\bq)
	& = - \frac{e}{V}  \sum_{\bm{k}} \frac{\hbar \bm{k}}{\me}
		c^{\dagger}_{\bm{k} - \frac{\bm{q}}{2}}
		c^{}_{\bm{k} + \frac{\bm{q}}{2}}
		+ e \sum_{\bq'} \bm{\alpha}_{\bq'} \times \bm{s} (\bq - \bq')
,\end{align}
where the first and second terms correspond to the currents of the normal and anomalous velocities, respectively.

 We denote the thermal Green function for the Hamiltonian $\mathcal{H}_0 + \mathcal{H}_{\rm R}$ as $G_{\bk, \bk'} (\zi \epsilon_n)$, which is evaluated up to the first order with respect to $\mathcal{H}_{\rm R}$ as
\begin{align}
G_{\bk, \bk'} (\zi \epsilon_n)
	 & \simeq g_{\bk} (\zi \epsilon_n) \delta_{\bk, \bk'}
	 + \frac{\hbar}{2} g_{\bk} (\zi \epsilon_n)
	 	\bm{\sigma} \cdot \bigl( \bm{\alpha}_{\bk - \bk'} \times (\bk + \bk') \bigr)
		g_{\bk'} (\zi \epsilon_n)
\label{eq:thermal_Green_function_H_0+H_R}
,\end{align}
where $\epsilon_n = (2 n + 1) \pi \kB T$ is the Matsubara frequency of fermion, $g_{\bk} (\zi \epsilon_n)$ is the thermal Green function for the Hamiltonian $\mathcal{H}_0$ given by
\begin{align}
g_{\bk} (\zi \epsilon_n)
	& = \frac{1}{\zi \epsilon_n - \epsilon_{\bk} + \zi \mathrm{sgn} (\epsilon_n) \hbar / (2 \tau)}
\label{eq:thermal_Green_function_H_0_delta-mu=0}
\end{align}
with the signum function $\mathrm{sgn} (x)$.
 The lifetime of electron evaluated within the Born approximation is given as
\begin{align}
\frac{\hbar}{2 \tau}
	& = \pi n_{\rm i} u^2 \nu
\label{eq:tau_def}
\end{align}
with $\nu = \nu (\eF)$ being the density of states~(DOS) at the Fermi energy $\eF$ of NM.

\section{\label{sec:non-local_current_by_magnetization_dynamics}non-local effective electric fields}
 By evaluating the non-local charge current induced by the magnetization dynamics, and using the Drude conductivity, we show that the charge current is driven by the non-local effective electric fields.
 We consider the exchange interaction up to the second order and the Rashba interaction in the first order in this section.

\subsection{\label{sec:sub:linear_response}Linear response to exchange interaction}
 For the linear response of the charge current $\braket{\bm{j} (\br, t)}^{(1)}$ to the external field $\bm{n} (\br', t')$, where the external Hamiltonian is given by $\mathcal{H}_{sd} (t')$, the current is calculated based on the Kubo formula~\cite{Kubo1991} (see Appendix~\ref{apx:sub:linear}) as
\begin{align}
\braket{j_{i} (\br, t)}^{(1)}
	& = \frac{\zi}{\hbar} \int^{t}_{-\infty} \mathrm{d}t' \braket{ [j_i (\br, t), \mathcal{H}_{sd} (t')] }
\notag \\
	& = \int \mathrm{d}\br' \int_{-\infty}^{\infty} \mathrm{d}t' \chi^{(1)}_{i j} (\br, \br'; t - t') M_{j} (\br', t')
\label{eq:j1}
,\end{align}
where $j_{i} (\br, t)$ is the Heisenberg representation of Eq.~(\ref{eq:current_density_def}), $[A, B] = AB - BA$ is the communicator, $\braket{\cdots}$ is the thermal average for $\mathcal{H}_0 + \mathcal{H}_{\rm R}$, and the linear response coefficient $\chi^{(1)}_{i j} (\br, \br'; t - t')$ is the retarded correction function between the charge current density and the spin density,
\begin{align}
\chi^{(1)}_{i j} (\br, \br'; t - t')
	& = - \frac{\zi}{\hbar} \theta (t - t') \braket{ [j_{i} (\br, t), s_j (\br', t')] }
\end{align}
with $\theta (t)$ being the Heaviside step function and $s_j (\br', t')$ being the Heisenberg representation of the spin density.
 Note that, since the Rashba field in the system has the spatial dependence, the linear response coefficient cannot be expressed as $\chi^{(1)}_{i j} (\br - \br'; t - t')$, which also means that the space translational symmetry is not assumed in the system.

 In the Fourier form, the charge current is given as
\begin{align}
\braket{j_{i} (\bq, \omega)}^{(1)}
	& = \sum_{\bq'} \chi^{\mathrm{R}, (1)}_{i j} (\bq, \bq'; \omega) M_{j} (\bq', \omega)
\end{align}
 Here, $\chi^{\mathrm{R}, (1)}_{i j} (\bq, \bq'; \omega)$ can be calculated from the following correlation function in the Matsubara formalism,
\begin{align}
\chi^{(1)}_{i j} (\bq, \bq' ; \zi \omega_{\lambda})
	& = - V \int_0^{\beta} \mathrm{d}\tau e^{\zi \omega_{\lambda} \tau}
		\braket{ \mathrm{T}_{\tau} j_{i} (\bq, \tau) s_{j} (- \bq', 0) }
\label{eq:chi_1_in_Matsubara_formalism}
,\end{align}
by taking the analytic continuation, $\zi \omega_{\lambda} \to \hbar \omega + \zi 0$ as 
\begin{align}
\chi^{\mathrm{R}, (1)}_{i j} (\bq, \bq'; \omega) = \chi^{(1)}_{i j} (\bq, \bq' ; \omega + \zi 0),
\end{align}
where $\beta = 1 / \kB T$ is the inverse temperature with the Boltzmann constant $\kB$, and $\omega_{\lambda} = 2 \pi \lambda  / \beta$ ($\lambda = 0, \pm 1, \cdots$) is the Matsubara frequency of boson.
 Note that the Matsubara frequencies are defined as in unit of energy instead of frequency.
\begin{figure}[hbtp]
	\centering
	\includegraphics[width=0.48\linewidth]{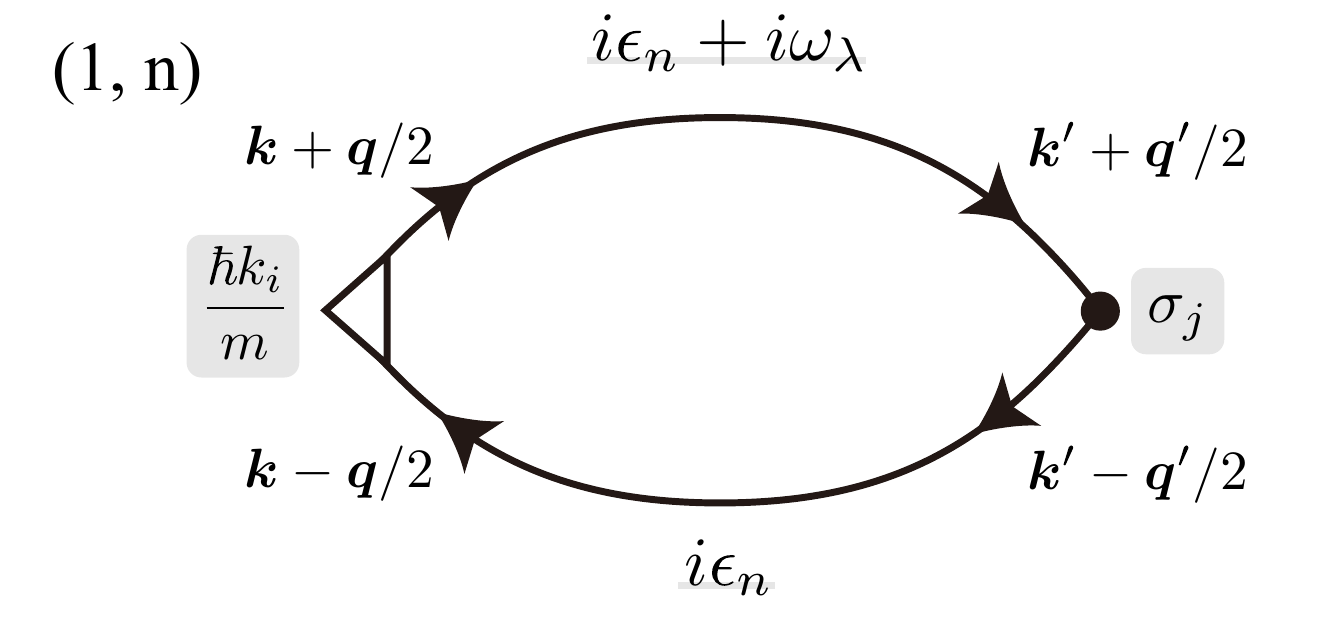}
	\includegraphics[width=0.48\linewidth]{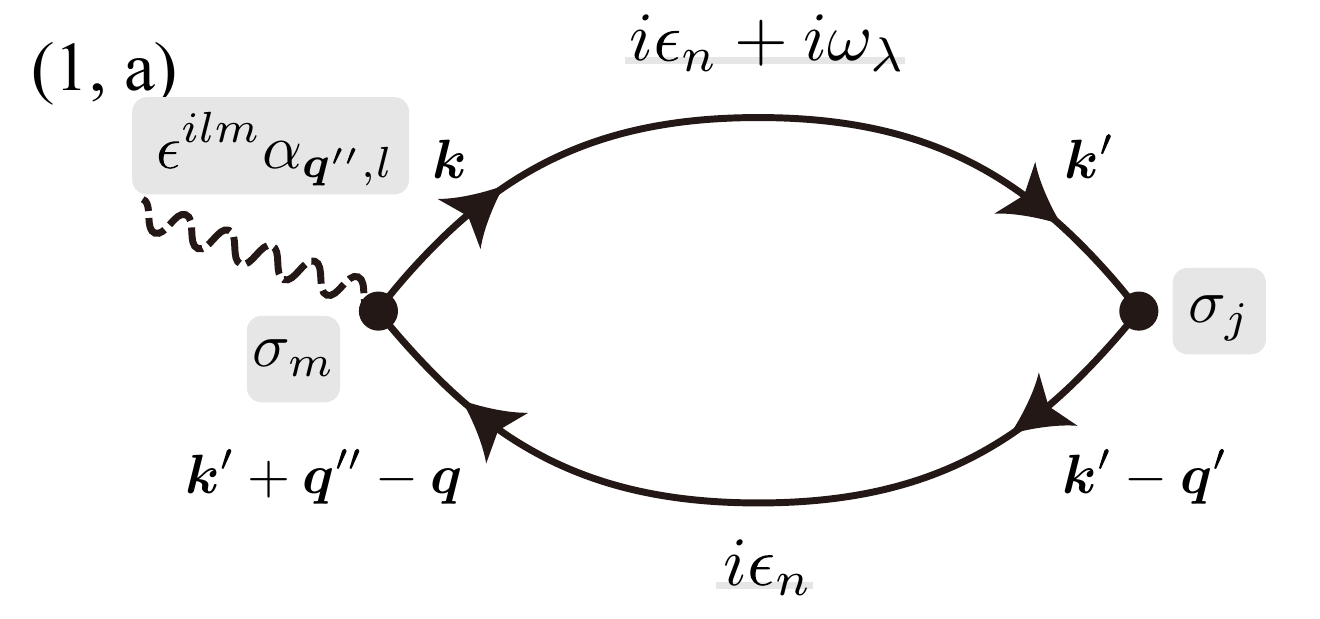}
	\caption{\label{fig:diagrams_chi_1n-1a}%
	The Feynman diagrams of $ \chi^{\rm (1, n)}_{i j}$ and $\chi^{\rm (1, a)}_{i j}$.
	The solid lines with two arrows denote the Green functions including the Rashba interaction given by Eq.~(\ref{eq:thermal_Green_function_H_0+H_R}), the filled circle represents the spin vertex, the unfilled triangle describes the normal velocity vertex, and the dashed wavy line indicates the anomalous velocity vertex without the Pauli matrix.
	}
\end{figure}
 By means of the thermal Green function for $\mathcal{H}_0 + \mathcal{H}_{\rm R}$, Eq.~(\ref{eq:chi_1_in_Matsubara_formalism}) is expressed as
\begin{align}
\chi^{(1)}_{i j} 
	& = \chi^{\rm (1, n)}_{i j}
		+ \chi^{\rm (1, a)}_{i j}
		+ \chi^{\rm (1, a)(df)}_{i j}
		+ \chi^{\rm (1, n)(df)}_{i j}
\label{eq:chi_1_to_3components}
,\end{align}
where $\chi^{\rm (1, n)}_{i j}$ and $\chi^{\rm (1, a)}_{i j}$ are the contributions from the normal and anomalous velocities without vertex corrections, which are given as
\begin{subequations}
\begin{align}
\chi^{\rm (1, n)}_{i j} (\bq, \bq'; \zi \omega_{\lambda})
	& = - \frac{e}{\beta V} \sum_{n} \sum_{\bk,\bk'} \frac{\hbar k_i}{\me} \mathrm{tr} \left[
		G_{\bk+\frac{\bq}{2}, \bk'+\frac{\bq'}{2}} (\zi \epsilon_n + \zi \omega_{\lambda})
		\sigma_j
		G_{\bk'-\frac{\bq'}{2}, \bk-\frac{\bq}{2}} (\zi \epsilon_n)
	\right]
\label{def:chi1n}
, \\
\chi^{\rm (1, a)}_{i j} (\bq, \bq'; \zi \omega_{\lambda})
	& = \frac{e}{\beta V} \sum_{n} \sum_{\bq''} \epsilon^{i l m} \alpha_{\bq'', l} \sum_{\bk, \bk'} \mathrm{tr} \left[
		\sigma_m 
		G_{\bk, \bk'} (\zi \epsilon_n + \zi \omega_{\lambda})
		\sigma_j
		G_{\bk'-\bq', \bk+\bq''-\bq} (\zi \epsilon_n)
	\right]
\label{def:chi1a}
\end{align}%
\label{eq:chi_1_with_thermal_Green_functions}
\end{subequations}
 Figure~\ref{fig:diagrams_chi_1n-1a} depicts $\chi^{\rm (1, n)}_{i j}$ and $\chi^{\rm (1, a)}_{i j}$.
 Contributions $\chi^{\rm (1, n)(df)}_{i j}$ and $\chi^{\rm (1, a)(df)}_{i j}$ contain the diffusion ladder VCs, whose diagrams and expressions are given in Appendix~\ref{apx:expressions} (Fig. \ref{fig:diagrams_chi_1}d-j).
 In order to evaluate them up to the first order of the Rashba interaction, we expand the Green functions in Eq.~(\ref{def:chi1n}) by using Eq.~(\ref{eq:thermal_Green_function_H_0+H_R}).
 As Eq.~(\ref{def:chi1a}) is already the first order of the Rashba interaction, the Green functions there can be approximated as $G_{\bk,\bk'} (\zi \epsilon_n) = g_{\bk} (\zi \epsilon_n) \delta_{\bk,\bk'}$.
 We take the analytic continuation, $\zi \omega_{\lambda}\to \hbar \omega + \zi 0$, and calculate the $\omega$-linear contribution, which leads to a contribution proportional to $\dot{\bm{M}}$.
 We also expand them up to the second order with respect to $\bq$ and $\bq'$.
 The details of the calculations are shown in Appendix~\ref{apx:diffusion_ladder_VC_calc}.
 Finally, we obtain 
\begin{align}
\chi^{\rm R, (1)}_{i j} (\bq, \bq'; \omega)
	& = e \epsilon^{m l j} \alpha_{\bq-\bq',l} \left(
		\eta^{(1)}_{\bq, \bq', i m}
		+ \zi \omega \, \varphi^{(1)}_{\bq, \bq', i m}
		+ \cdots
	\right)
\label{eq:res:chi1}
, \\
\varphi^{(1)}_{\bq, \bq', i m}
	& = \frac{2 \nu \tau}{q'^2} \left( \delta_{i m} (\bq - \bq') \cdot \bq' - (q_i - q'_i) q'_m \right)
,\end{align}
where $\eta^{(1)}_{\bq, \bq', i m}$ is the static response to the magnetization, and $\varphi^{(1)}_{\bq, \bq', i m}$ is the dynamical response of our interest.
 In the real space, using the Drude conductivity $\sigma_{D} = 2 e^2 \nu \eF \tau / (3 \me) = e^2 \nu D_0$ with $\nu = \nu (\eF)$ being DOS at the Fermi energy of NM and $D_0$ being the diffusion constant, we find the linear-order current as 
\begin{align}
\braket{\bm{j} (\br, t) }^{(1)}
	& =  - \frac{e l^2 \nu \tau}{3} \int {\mathrm{d}\br'} D(\bm{r}-\bm{r}')
		\Bigl(
			\bigl( \bm{\nabla}_{\br'} \cdot \bm{\nabla}_{\br} \bigr) [\bm{\alpha} (\br) \times \dot{{\bm{M}}} (\br', t)]
			- \bm{\nabla}_{\br} \bigl(\bm{\nabla}_{\br'} \cdot [\bm{\alpha} (\br) \times \dot{{\bm{M}}} (\br', t)] \bigr)
		\Bigr)
\\	& =  - \frac{e l^2 \nu \tau}{3} \int {\mathrm{d}\br'} D(\bm{r}-\bm{r}')
		\biggl[\bm{\nabla}_{\br'} \times \Bigl( \bm{\nabla}_{\br} \times  [\bm{\alpha} (\br) \times \dot{{\bm{M}}} (\br', t)]  \Bigr)\biggr],
\end{align}
where 
\begin{align} 
D(\bm{r}) \equiv \frac{1}{V} \sum_{\bm{q}} \frac{e^{i\bm{q}\cdot\bm{r}}}{D_0 q^2 \tau}=\frac{3}{4\pi l^2}\frac{1}{r},
\end{align}
is  the diffusion propagator, $l\equiv \sqrt{3D_0\tau}$ being the elastic mean free path.
%

%
%
%
%
%
\subsection{\label{sec:sub:second_order_response}Second order response to exchange interaction}
 For the second order response to the exchange Hamiltonian, the charge current is given by
\begin{align}
\braket{j_i (\br, t)}^{(2)}
	& = \left(\frac{\zi}{\hbar}\right)^2 \int_{-\infty}^{t} \mathrm{d}t' \int_{-\infty}^{t'} \mathrm{d}t''
		\braket{ \Bigl[
				[j_i (\br, t), \mathcal{H}_{sd} (t')], \mathcal{H}_{sd} (t'')
		\Bigr] }
\notag \\
	& = \iint \mathrm{d}\br' \mathrm{d}\br'' \iint_{-\infty}^{\infty} \mathrm{d}t' \mathrm{d}t''
		\chi^{\rm R, (2)}_{i j k} (\br, \br', \br''; t - t', t' - t'')
		M_j (\br', t') M_k (\br'', t'')
\label{eq:j2}
,\end{align}
where the second order response coefficient $\chi^{\rm R, (2)}_{i j k} (\br, \br', \br''; t - t', t' - t'')$ is expressed as
\begin{align}
\chi^{\rm R, (2)}_{i j k} (\br, \br', \br''; t - t', t' - t'')
	& = \frac{1}{2} \Bigl(
			Q_{i j k} (\br, \br', \br''; t - t', t' - t'')
			+ Q_{i k j} (\br, \br'', \br'; t - t'', t'' - t')
		\Bigr)
, \\
Q_{i j k} (\br, \br', \br''; t - t', t' - t'')
	& = \left(\frac{\zi}{\hbar}\right)^2
		\theta (t - t') \theta (t' - t'') \braket{ \Bigl[ [j_{i} (\br, t), s_j (\br', t')], s_k (\br'', t'') \Bigr] }
\end{align}
 Here, $\chi^{\rm R, (2)}_{i k j} (\br, \br'', \br'; t - t'', t'' - t') = \chi^{\rm R, (2)}_{i j k} (\br, \br', \br''; t - t', t' - t'')$.
 In the Fourier form, the current is given as
\begin{align}
\braket{j_i (\bq, \omega)}^{(2)}
	& = \sum_{\bq',\bq''} \int_{-\infty}^{\infty} \frac{\mathrm{d} \omega'}{2 \pi}
		\chi^{\rm R, (2)}_{i j k} (\bq, \bq', \bq''; \omega, \omega')
		M_j (\bq', \omega - \omega') M_k (\bq'', \omega')
\end{align}
 From Appendix~\ref{apx:sub:second}, the non-linear response coefficient $\chi^{\rm R, (2)}_{i j k} (\bq, \bq', \bq''; \omega, \omega')$ is evaluated from
\begin{align}
\chi^{(2)}_{i j k} (\bq, \bq', \bq'' ;\zi \omega_{\lambda}, \zi \omega_{\lambda'})
	& = \frac{V^2}{2} \iint_{0}^{\beta} \mathrm{d}\tau \mathrm{d}\tau' e^{\zi \omega_{\lambda} \tau + \zi \omega_{\lambda'} \tau'}
		\braket{ \mathrm{T}_{\tau} j_{i} (\bq, \tau + \tau') s_{j} (- \bq', \tau') s_{k} (- \bq'', 0) }
\label{eq:chi_2_in_Matsubara_formalism}
\end{align}
by taking the analytic continuations as
\begin{align}
\zi \omega_{\lambda} \to \hbar \omega + 2 \zi 0
, \quad
\zi \omega_{\lambda'} \to \hbar \omega' + \zi 0
\label{eq:analytic_continuation}
\end{align}
 Note that, in order to obtain the precise response coefficient, the order of the analytic continuations for $\zi \omega_{\lambda}$ and $\zi \omega_{\lambda'}$ must be specified; taking $\zi \omega_{\lambda'}$ to the real frequency $\hbar \omega'$ and then taking $\zi \omega_{\lambda}$ to $\hbar \omega$ from the upper plane ($\omega_{\lambda^{(\prime)}} > 0$).
 Hence, we have set Eq.~(\ref{eq:analytic_continuation}).
 We should emphasise that the Matsubara Green function method can apply to the non-linear responses as demonstrated by Jujo~\cite{Jujo2006} and by Kohno and Shibata~\cite{Kohno2007}. (See Appendix~\ref{apx:nonlinear_response_theory} for general cases.)

\begin{figure}[hbtp]
	\centering
	\includegraphics[width=0.9\linewidth]{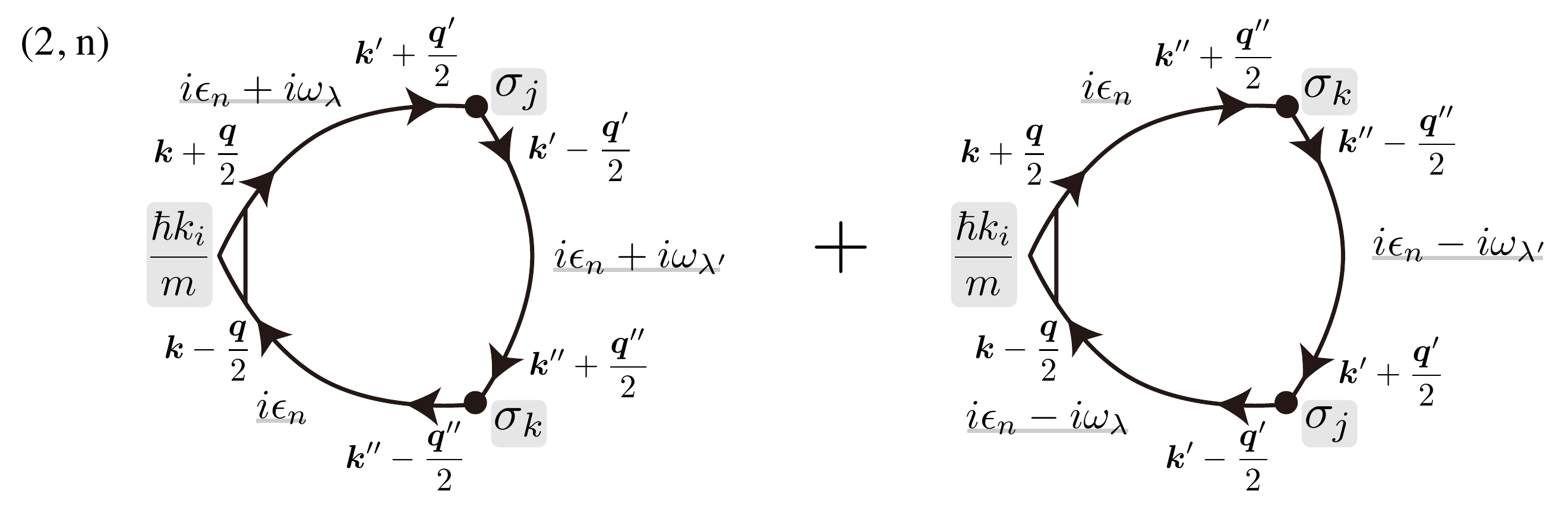}
	\includegraphics[width=0.9\linewidth]{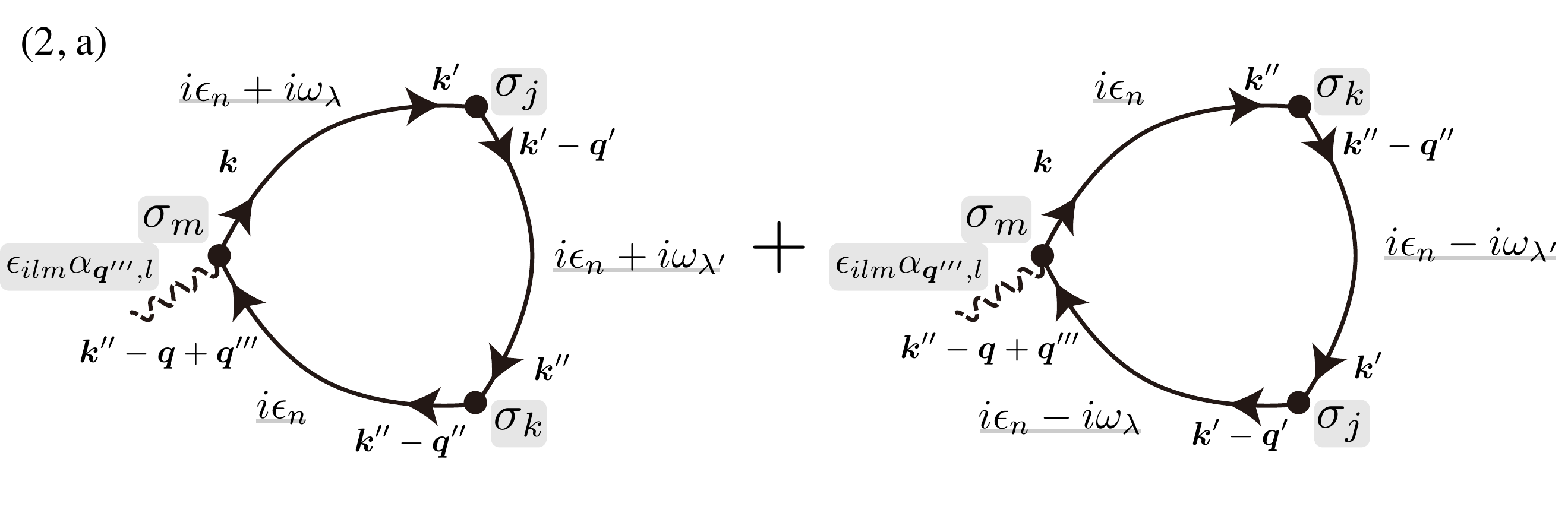}
	\caption{\label{fig:diagrams_chi_2n-2a}%
	The Feynman diagrams of $ \chi^{\rm (2, n)}_{i j k}$ and $\chi^{\rm (2, a)}_{i j k}$.
	The lines and symbols are defined in the caption of Fig.~\ref{fig:diagrams_chi_1n-1a}.
	}
\end{figure}
 Here, we separate Eq.~(\ref{eq:chi_2_in_Matsubara_formalism}) into three components in the similar way of the calculation of $\chi^{(1)}_{ij}$ [Eq.~(\ref{eq:chi_1_to_3components})],
\begin{align}
\chi^{(2)}_{i j k}
	= \chi^{\rm (2,n)}_{i j k}
	+ \chi^{\rm (2,a)}_{i j k}
	+ \chi^{\rm (2,n)(df)}_{i j k}
	+ \chi^{\rm (2,a)(df)}_{i j k}
\label{eq:chi_2_to_3components}
,\end{align}
where the first two terms are the contributions from the normal and anomalous velocities without VCs, respectively, given by
\begin{align}
\chi^{\rm (2,n)}_{i j k} (\bq, \bq', \bq''; \zi \omega_{\lambda}, \zi \omega_{\lambda'})
	& = - \frac{e}{2 \beta V} \sum_{n} \sum_{\bk,\bk',\bk''} \frac{\hbar k_i}{m}
\notag \\ & \hspace{-3em} \times
		\mathrm{tr} \Bigl[
			G_{\bk+\frac{\bq}{2},\bk'+\frac{\bq'}{2}} (\zi \epsilon_n + \zi \omega_{\lambda})
			\sigma_j G_{\bk'-\frac{\bq'}{2}, \bk''+\frac{\bq''}{2}} (\zi \epsilon_n + \zi \omega_{\lambda'})
			\sigma_k G_{\bk''-\frac{\bq''}{2}, \bk-\frac{\bq}{2}} (\zi \epsilon_n)
\notag \\ & \hspace{-2em}
			+ G_{\bk+\frac{\bq}{2},\bk''+\frac{\bq''}{2}} (\zi \epsilon_n)
			\sigma_k G_{\bk''-\frac{\bq''}{2}, \bk'+\frac{\bq'}{2}} (\zi \epsilon_n - \zi \omega_{\lambda'})
			\sigma_j G_{\bk'-\frac{\bq'}{2}, \bk-\frac{\bq}{2}} (\zi \epsilon_n - \zi \omega_{\lambda})
		\Bigr]
\label{eq:chi_2n_def}
, \\
\chi^{\rm (2,a)}_{i j k} (\bq, \bq', \bq''; \zi \omega_{\lambda}, \zi \omega_{\lambda'})
	& = \frac{e}{2 \beta V} \sum_{n} \sum_{\bk,\bk',\bk''} \epsilon_{i l m} \sum_{\bq'''} \alpha_{\bq''',l}
\notag \\ & \hspace{-3em} \times
		\mathrm{tr} \Bigl[
			\sigma_m G_{\bk,\bk'} (\zi \epsilon_n + \zi \omega_{\lambda})
			\sigma_j G_{\bk'-\bq', \bk''} (\zi \epsilon_n + \zi \omega_{\lambda'})
			\sigma_k G_{\bk''-\bq'', \bk-\bq+\bq'''} (\zi \epsilon_n)
\notag \\ & \hspace{-2em}
			+ \sigma_m G_{\bk,\bk''} (\zi \epsilon_n)
			\sigma_k G_{\bk''-\bq'', \bk'} (\zi \epsilon_n - \zi \omega_{\lambda'})
			\sigma_j G_{\bk'-\bq', \bk-\bq+\bq'''} (\zi \epsilon_n - \zi \omega_{\lambda})
		\Bigr]
\label{eq:chi_2a_def}
\end{align}
 The last two terms in Eq.~(\ref{eq:chi_2_to_3components}) include the ladder type VCs of $\chi^{\rm (2,n)}_{i j k}$ and $\chi^{\rm (2,a)}_{i j k}$, which are given in Appendix~\ref{apx:expressions}.
 We evaluate them up to the first order of the Rashba interaction.
 For $\chi^{\rm (2,n)}_{i j k}$, we expand the Green functions in the first term of Eq.~(\ref{eq:chi_1_with_thermal_Green_functions}) by using Eq.~(\ref{eq:thermal_Green_function_H_0+H_R}).
 As $\chi^{\rm (2,a)}_{i j k}$ is already the first order of the Rashba interaction, the Green functions in the second term can be approximated as $G_{\bk, \bk'} (\zi \epsilon_n) = g_{\bk} (\zi \epsilon_n) \delta_{\bk,\bk'}$.
 We take the analytic continuation as shown by Eq.~(\ref{eq:analytic_continuation}) and evaluate the $\omega'$-linear contribution, which leads to a contribution proportional to $\bm{M} \times \dot{\bm{M}}$.
 Then, we expand them up to the second order with respect to $\bq$, $\bq'$ and $\bq''$.
 The details of the calculations are shown in Appendix~\ref{apx:diffusion_ladder_VC_calc}.
 After all, we have the followings:
\begin{align}
\chi^{\rm R, (2)}_{i j k} (\bq, \bq', \bq''; \omega, \omega')
	& = 2 \zi e \epsilon^{o l m} \epsilon^{m j k} \alpha_{\bq-\bq'-\bq'',l}
		\left(
		\eta^{(2)}_{\bq, \bq', \bq'', i o}
		+ \zi \omega \, \vartheta^{(2)}_{\bq, \bq', \bq'', i o}
		+ \zi \omega' \, \varphi^{(2)}_{\bq, \bq', \bq'', i o}
		+ \cdots
		\right)
, \\
\varphi^{(2)}_{\bq, \bq', \bq'', i o}
	& = - \frac{2 \zi \nu \tau^2}{\hbar} \frac{\delta_{i o} (\bq - \bq' - \bq'') \cdot (\bq' + \bq'') - (q_i - q'_i - q''_i) (q'_o + q''_o) }{(\bq' + \bq'')^2}
,\end{align}
where $\eta^{(2)}_{\bq, \bq', \bq'', i o}$ is static response to the magnetization, $\vartheta^{(2)}_{\bq, \bq', \bq'', i o}$ is the dynamical response proportional to $\mathrm{d} (M_j M_k) / \mathrm{d}t$, which are negligible.
$\varphi^{(2)}_{\bq, \bq', \bq'', i o}$ is the dynamical response of our interest, which is proportional to such as $M_j \mathrm{d} (M_k) / \mathrm{d}t$.
 In the real space, dynamically-induced current is 
\begin{align}
\braket{ \bm{j}(\br, t)}^{(2)} 
	& = \frac{4 e l^2 \nu \tau^2}{3 \hbar} \int {\mathrm{d}\br'} D( \br - \br')
		\biggl[ \bm{\nabla}_{\br'} \times \biggl( \bm{\nabla}_{\br} \times 
		\biggl[ \bm{\alpha} (\br) \times \biggl( {\bm{M}} (\br', t) \times \dot{{\bm{M}}} (\br', t) \biggr)
 \biggr] \biggr)\biggr]
 \label{eq:res:chi2}
.\end{align}

\section{\label{sec:results_discussion} Results and Discussion}
The generated charge current to the second order responses to the exchange interaction is therefore $\bm{j}(\br, t)= \braket{ \bm{j}(\br, t)}^{(1)} + \braket{ \bm{j}(\br, t)}^{(2)} $.
The current is expressed as a response to a non-local effective electric field as 
$\bm{j} (\br, t) = \sigma_{D} \bm{E}_{\rm eff} (\br, t)$, where 
\begin{align}
\bm{E}_{\rm eff} (\br, t)
	& = \frac{\me l^2 }{2 e \eF} \int {\mathrm{d}\br'} D( \br - \br')
		\biggl[ \bm{\nabla}_{\br'} \times \biggl( \bm{\nabla}_{\br} \times 
		\biggl[ \bm{\alpha} (\br) \times 
		\biggl( \dot{{\bm{M}}} (\br', t)+\frac{4\tau}{\hbar}[ {\bm{M}} (\br', t) \times \dot{{\bm{M}}} (\br', t)] \biggr)
 \biggr] \biggr)\biggr]
\label{eq:res_case1}
.\end{align}
Note that the magnetization $\bm{M} (\br, t)$ is defined as including the $sd$ interaction strength.
The linear response term, $\bm{E}_{\rm eff}^{(1)}$, is written  as $\bm{E}_{\rm eff}^{(1)}=-\dot{\bm{A}}_{\rm eff}^{(1)}$, where 
\begin{align}
\bm{A}_{\rm eff}^{(1)} (\br, t)
	& = -\frac{\me l^2}{2 e \eF} \int {\mathrm{d}\br'} D( \br - \br')
		\biggl[ \bm{\nabla}_{\br'} \times \biggl( \bm{\nabla}_{\br} \times 
		\biggl[ \bm{\alpha} (\br) \times 
		 {{\bm{M}}} (\br', t)
 \biggr] \biggr)\biggr]
\label{eq:res_case11}
\end{align}
is a non-local extension of effective gauge field discussed in Refs. \cite{Kim2012,Takeuchi2012,Nakabayashi2014}.
In contrast, the second-order contribution, proportional to spin damping, $\bm{M} \times {\dot{\bm{M}}}$, does not have the corresponding  gauge field like in the local case \cite{Tatara_smf13}.

For junctions like a trilayer homogeneous in the $xy$-plane, the spatial derivative is finite only in the $z$-direction. 
The in-plane current, which is of experimental interest, in this case reads 
\begin{align}
\bm{j}^{\parallel} (\br, t)
	& = \frac{\me l^2}{2 e \eF}\sigma_{ D} \int {\mathrm{d}\br'} D( \br - \br')
		(\bm{\nabla}_{\br}^z\bm{\alpha} (\br)) \times \bm{\nabla}_{\br'}^z
		\biggl[ \dot{{\bm{M}}} (\br', t)+\frac{4\tau}{\hbar}[ {\bm{M}} (\br', t) \times \dot{{\bm{M}}} (\br', t)] \biggr]
\label{eq:res_casepara}
.\end{align}
This result indicates that the spatially-inhomogeneity of precessing spin at the F-NM interface drives an in-plane effective motive force at the NM-HM interface as a result of electron diffusion. This motive force is an alternative and direct interpretation of inverse Rashba-Edelstein effect.

For describing the case of a spacer  thicker than spin diffusion length, spin relaxation effect needs to be included in the diffusion.
As was discussed in Ref.  \cite{Tatara18}, the result in this case  becomes  Eq. (\ref{eq:res_case1}) with diffusion $D(\br)$ replaced by the one including spin diffusion length,
\begin{align} 
D_s(\bm{r}) \equiv \frac{1}{V} \sum_{\bm{q}} \frac{e^{i\bm{q}\cdot\bm{r}}}{D_0 q^2 \tau+\gamma_s },
\end{align}
where $\gamma_s$, proportional to spin relaxation rate, is related to spin diffusion length 
$l_{s}$ as $l_{s}=l/\sqrt{3\gamma_s}$.

The non-local effective electric field found in the present study is an electric counterpart of the non-local effective magnetic field (non-local spin Berry's phase) discussed in the context of the anomalous Hall effect \cite{Nakazawa2014}. 
Although the spin Berry's phase itself arises from static magnetization textures, calculation of the non-local contribution in the present formalism requires including an external field with a finite or infinitesimal frequency, as the electron diffusion applies to non-equilibrium situations only. 

\begin{acknowledgments}
 JF would like to thank A.~Shitade and S.C.~Furuya for giving informative comments.
JF is supported by a Grant-in-Aid for Specially Promoted Research (No.~15H05702).
GT thanks 
a Grant-in-Aid for Exploratory Research (No.16K13853) and a Grant-in-Aid for Scientific Research (B) (No. 17H02929) from the Japan Society for the Promotion of Science,
a Grant-in-Aid for Scientific Research on Innovative Areas (No.26103006) from The Ministry of Education, Culture, Sports, Science and Technology (MEXT), Japan, 
and 
the Graduate School Materials Science in Mainz (MAINZ) (DFG GSC 266) for finantial support.
\end{acknowledgments}

\appendix

\section{\label{apx:nonlinear_response_theory}Matsubara formalism for non-linear responses}
 Here, we show the formulation of the non-linear response theory based on the Matsubara formalism.
 Some textbooks~\cite{Zagoskin1998,Rickayzen2013} explain that it is not possible.
 However, by a careful treatment of analytic continuations, this formulation can be done and leads to the exactly same result from the Keldysh formalism.
 In this Appendix, we show the way to evaluate the responses up to the second order with respect to the external force.
 We also discuss briefly the evaluations for the higher order responses.

\subsection{Setup}
 In this Appendix, we assume that the system we consider is expressed by the Hamiltonian $\mathcal{H}$, and the mechanical external force is $F_{\mu} (t)$ ($\mu$ is index), which couples to the physical quantity $\hat{A}_{\mu}$, hence the external Hamiltonian given by
\begin{align}
\hat{\mathcal{H}}' (t)
	& = - \hat{A}_{\mu} F_{\mu} (t)
\end{align}
 (One should presume that the dummy index $\mu$ sums over all the external forces.)
 We introduce $\eta > 0$ as an infinitesimal quantity to ensure that the system is in thermal equilibrium and the external force is zero at the time $t \to - \infty$, and the external force is turned on adiabatically from the time:
\begin{align}
F_{\mu} (t)
	& = e^{\eta t} \int_{-\infty}^{\infty} \frac{\mathrm{d}\omega}{2 \pi} e^{- \zi \omega t} F_{\mu} (\omega)
	= \int_{-\infty}^{\infty} \frac{\mathrm{d}\omega}{2 \pi} e^{- \zi (\omega + \zi \eta) t} F_{\mu} (\omega)
\label{eq:convergence_factor}
\end{align}
 Following the paper by Kubo~\cite{Kubo1991}, the response of the physical quantity $\hat{B}$ to the external force $F_{\mu}$ is given by $\langle\hat{B}\rangle (t) = \langle\hat{B}\rangle_0 + \varDelta_{1} B (t) + \varDelta_{2} B (t) + \cdots + \varDelta_{k} B (t) + \cdots$ with the $k$-th order response
\begin{align}
\varDelta_{k} B (t)
	& = \left( \frac{-1}{\zi \hbar} \right)^k \int_{-\infty}^{t} \mathrm{d}t_1 \int_{-\infty}^{t_1} \mathrm{d}t_2 \cdots \int_{-\infty}^{t_{k-1}} \mathrm{d}t_k
\notag \\[1ex] & \hspace{3em} \times
		\mathrm{Tr} \left\{
			\bigg[ \hat{A}_{\mu_1} (t_1), \Big[ \hat{A}_{\mu_2} (t_2), \big[ \cdots , [ \hat{A}_{\mu_k} (t_k), \hat{\rho} ] \big] \cdots \Big] \bigg]
			\hat{B} (t)
		\right\}
		F_{\mu_1} (t_1) F_{\mu_2} (t_2) \cdots F_{\mu_k} (t_k)
\label{eq:apx:Kubo_formula}
,\end{align}
where $\langle\hat{B}\rangle_0$ is the expectation value without any external fields, $\hat{A} (t) = e^{\zi \hat{\mathcal{H}} t / \hbar} \hat{A} e^{- \zi \hat{\mathcal{H}} t / \hbar}$ is the Heisenberg representation of $\hat{A}$, $[\hat{A}, \hat{B}] = \hat{A} \hat{B} - \hat{B} \hat{A}$ is communicator, $\hat{\rho}$ is the density matrix operator for $\hat{\mathcal{H}}$, and
\begin{align}
\hat{\rho}
	& = e^{- \beta \hat{\mathcal{H}}} / \mathrm{Tr\,} e^{- \beta \hat{\mathcal{H}}}
	= e^{\beta ( \Omega - \hat{\mathcal{H}} ) }
\end{align}
with $\beta = 1 / \kB T$ and with $\Omega = - \kB T \ln \mathrm{Tr} \,\{ e^{- \beta \hat{\mathcal{H}}} \}$ being the thermodynamic potential.
 Introducing $\ket{n}$ as the eigenstates of the Hamiltonian, $\hat{\mathcal{H}} \ket{n} = E_n \ket{n}$, $\mathrm{Tr} \,\{ \,\cdots \}$ is given as
\begin{align}
\mathrm{Tr}\,\{ \hat{A} \}
	& = \sum_{n} \bra{n} \hat{A} \ket{n}
\label{eq:expansion_by_eigenstates}
\end{align}
 The thermal average $\braket{\cdots}$ for the system $\mathcal{H}$ in the temperature $T$ is defined by
\begin{align}
\braket{ \hat{A} }
	& = \mathrm{Tr}\,\{ \hat{\rho} \hat{A} \}
	= \sum_{n} e^{\beta (\Omega - E_n)} \bra{n} \hat{A} \ket{n}
\label{eq:thermal_average}
\end{align}
 We also note that the time translational symmetry is held in thermal equilibrium.

\subsection{\label{apx:sub:linear}Linear response}
 We first look at the linear response ($k = 1$).
 Using the cyclic relation, $\mathrm{Tr}\,\{ \hat{A} \hat{B} \hat{C} \} = \mathrm{Tr}\,\{ \hat{B} \hat{C} \hat{A} \} = \mathrm{Tr}\,\{ \hat{C} \hat{A} \hat{B} \}$, we find
\begin{align}
\varDelta_1 B (t)
	& = \frac{-1}{\zi \hbar} \int_{-\infty}^{t} \mathrm{d}t'
	\mathrm{Tr} \left\{
		[ \hat{A}_{\mu} (t'), \hat{\rho} ] \hat{B} (t)
	\right\}
	F_{\mu} (t')
\notag \\
	& = \int_{-\infty}^{\infty} \mathrm{d}t' Q_{\mu}^{\mathrm{R}} (t - t') F_{\mu} (t')
,\end{align}
where $Q_{\mu}^{\mathrm{R}} (t)$ is the retarded two-point Green function,
\begin{align}
Q_{\mu}^{\mathrm{R}} (t)
	& = \frac{\zi}{\hbar} \theta (t) \braket{ [\hat{B} (t), \hat{A}_{\mu} (0)] }
\end{align}
 By using the Fourier transformation\footnote{As we have introduced the convergence factor $\eta$ as in Eq.~(\ref{eq:convergence_factor}), $\varDelta_1 B (t)$ is also assumed to be expressed as $\varDelta_1 B (t) = e^{\eta t} \int \varDelta_1 B (\omega) e^{- \zi \omega t} \mathrm{d}\omega/2 \pi$ for the time-translational symmetry.}, $\varDelta_1 B (\omega) = Q_{\mu}^{\mathrm{R}} (\omega) F_{\mu} (\omega)$, where
\begin{align}
Q_{\mu}^{\mathrm{R}} (\omega)
	& = \frac{\zi}{\hbar} \int_{0}^{\infty} \mathrm{d}t e^{ \zi (\omega + \zi 0 ) t } \braket{ [\hat{B} (t), \hat{A}_{\mu} (0)] }
\end{align}
 Here, $\omega + \zi 0$ stands for $\lim_{\eta \to 0+} \omega + \zi \eta$.

 The Matsubara Green function corresponding to $Q_{\mu}^{\mathrm{R}} (\omega)$ is 
\begin{align}
\mathcal{Q}_{\mu} (\zi \omega_{\lambda})
	& = \frac{1}{\hbar} \int_{0}^{\hbar\beta} \mathrm{d}\tau e^{\zi \omega_{\lambda} \tau}
	\braket{ \mathrm{T}_{\tau} \{ \hat{B} (\tau) \hat{A}_{\mu} (0) \} }
,\end{align}
where $\omega_{\lambda} = 2 \pi \lambda / \hbar \beta$ $( \lambda = 0, \pm 1, \pm 2, \cdots )$ is the Matsubara frequency of bosons, $\tau$ is the imaginary time, and $\hat{A} (\tau) = e^{\hat{\mathcal{H}} \tau / \hbar} \hat{A} e^{-\hat{\mathcal{H}} \tau / \hbar}$ is the so-called Heisenberg representation of $\hat{A}$ in the imaginary time ($t = - \zi \tau$) and defined in the region $- \hbar \beta \le \tau \le \hbar\beta$.
 $\mathrm{T}_{\tau} \{ \cdots \}$ is the time ordering operator of $\tau$.
 Note that as one shows the periodicity $\langle \mathrm{T}_{\tau} \{ \hat{B} (\tau - \hbar \beta) \hat{A}_{\mu} (0) \} \rangle = \langle \mathrm{T}_{\tau} \{ \hat{B} (\tau) \hat{A}_{\mu} (0) \} \rangle$ for $\tau \ge 0$ using Eq.~(\ref{eq:thermal_average}), it can be expressed by means of the Fourier series of $e^{\zi \omega_{\lambda} \tau}$.

 The correspondence between $Q_{\mu}^{\mathrm{R}} (\omega)$ and $\mathcal{Q}_{\mu} (\zi \omega_{\lambda})$ is proven easily by representing them in the Lehmann representation and taking the analytic continuation, $\zi \omega_{\lambda} \to \omega + \zi 0$, resulting to
\begin{align}
Q_{\mu}^{\mathrm{R}} (\omega)
	& = \mathcal{Q}_{\mu} (\omega + \zi 0)
\end{align}
 From these, we can evaluate the linear response coefficient $Q_{\mu}^{\mathrm{R}} (\omega)$ from the corresponding Matsubara Green function $\mathcal{Q}_{\mu} (\zi \omega_{\lambda})$ by taking the analytic continuation, $\zi \omega_l \to \omega + \zi 0$.

\subsection{\label{apx:sub:second}Second order response}
 Next, we show the way to evaluate the second order response precisely.
 This procedure is similar to the evaluation of the linear response; (1) find the correlation function in the Matsubara formalism corresponding to the response coefficient, (2) calculate the correlation function, and (3) take the precise analytic continuation.
 The procedures (1) and (3) are of the central theme in this Appendix since the procedure (2) is same as the well-known procedure.

 For $k = 2$ in Eq.~(\ref{eq:apx:Kubo_formula}), the second order response is given as
\begin{align}
\varDelta_2 B (t)
	& = \left( \frac{-1}{\zi \hbar} \right)^2
		\int_{-\infty}^{t} \mathrm{d}t_1 \int_{-\infty}^{t_1} \mathrm{d}t_2
		\mathrm{Tr}\left\{
			\Bigl[ \hat{A}_{\mu} (t_1), \bigl[ \hat{A}_{\nu} (t_2), \hat{\rho} \bigr] \Bigl] \hat{B} (t)
		\right\}
		F_{\mu} (t_1) F_{\nu} (t_2)
\notag \\
	& = \int_{-\infty}^{\infty} \mathrm{d}t_1 \int_{-\infty}^{\infty} \mathrm{d}t_2
		Q_{\mu \nu}^{\mathrm{R}} (t, t_1, t_2) F_{\mu} (t_1) F_{\nu} (t_2)
\notag \\
	& = \frac{1}{2!} \int_{-\infty}^{\infty} \mathrm{d}t_1 \int_{-\infty}^{\infty} \mathrm{d}t_2
		\Bigl( Q_{\mu \nu}^{\mathrm{R}} (t, t_1, t_2) + Q_{\nu \mu}^{\mathrm{R}} (t, t_2, t_1) \Bigr) F_{\mu} (t_1) F_{\nu} (t_2)
\label{eq:2nd_response_formula}
,\end{align}
where $Q_{\mu \nu}^{\mathrm{R}} (t, t', t'')$ is the retarded three-points correlation function given by
\begin{align}
Q_{\mu \nu}^{\mathrm{R}} (t, t', t'')
	& = - \frac{1}{\hbar^2} \theta (t - t') \theta (t' - t'')
		\braket{ \bigl[ [ \hat{B} (t), \hat{A}_{\mu} (t') ], \hat{A}_{\nu} (t'') \bigr] }
\end{align}
 Here, we should point out that we treated the external forces $F_{\mu} (t_1)$ and $F_{\nu} (t_2)$ symmetrically; we added the term interchanging $\mu$ and $\nu$ as well as $t_1$ and $t_2$ and divided them by $2!$ as in the last equal of Eq.~(\ref{eq:2nd_response_formula}).
 From the following relation by using Eq.~(\ref{eq:expansion_by_eigenstates}),
\begin{align}
Q_{\mu \nu}^{\mathrm{R}} (t, t_1, t_2)
	& = - \frac{1}{\hbar^2} \theta (t - t_1) \theta (t_1 - t_2)
		\sum_{n,m,l} e^{\beta (\Omega - E_n)} ( 1 - e^{\beta ( E_n - E_l )} )
\notag \\ & \hspace{1em} \times \left\{
		e^{\zi (E_n - E_m) (t - t_1)/\hbar + \zi (E_n - E_l) (t_1 - t_2)/\hbar }
		\bra{n} \hat{B} \ket{m}
		\bra{m} \hat{A}_{\mu} \ket{l}
		\bra{l} \hat{A}_{\nu} \ket{n}
		+ \mathrm{(c.c.)}
	\right\}
,\end{align}
one find $Q_{\mu \nu}^{\mathrm{R}} (t, t_1, t_2) = Q_{\mu \nu}^{\mathrm{R}} (t - t_1, t_1 - t_2)$ with
\begin{align}
Q_{\mu \nu}^{\mathrm{R}} (t, t')
	& = - \frac{1}{\hbar^2} \theta (t) \theta (t')
		\braket{ \bigl[ [ \hat{B} (t + t'), \hat{A}_{\mu} (t') ], \hat{A}_{\nu} (0) \bigr] }
\end{align}
 By means of $Q_{\mu \nu}^{\mathrm{R}} (t, t')$, Eq.~(\ref{eq:2nd_response_formula}) is rewritten as
\begin{align}
\varDelta_2 B (t)
	& = \int_{-\infty}^{\infty} \mathrm{d}t_1 \int_{-\infty}^{\infty} \mathrm{d}t_2
		\phi^{\mathrm{R}}_{\mu \nu} (t - t_1, t_1 - t_2)
		F_{\mu} (t_1) F_{\nu} (t_2)
, \\
\phi^{\mathrm{R}}_{\mu \nu} (t, t')
	& = \frac{1}{2!} \Bigl( Q_{\mu \nu}^{\mathrm{R}} (t, t') + Q_{\nu \mu}^{\mathrm{R}} (t + t', - t') \Bigr)
\end{align}
 Then, the second order response [Eq.~(\ref{eq:2nd_response_formula})] is expressed in the Fourier space\footnote{Considering the convergence factors, we have assumed $\varDelta_2 B (t) = e^{(\eta + \eta') t} \int \varDelta_2 B (\omega) e^{-\zi \omega t} \mathrm{d}t/2 \pi$}  as
\begin{align}
\varDelta_2 B (\omega)
	& = \int_{-\infty}^{\infty} \frac{\mathrm{d}\omega'}{2 \pi}
		\phi^{\mathrm{R}}_{\mu \nu} (\omega, \omega')
		F_{\mu} (\omega - \omega') F_{\nu} (\omega')
\label{eq:second_order_response}
, \\
\phi^{\mathrm{R}}_{\mu \nu} (\omega, \omega')
	& = ( Q_{\mu \nu}^{\mathrm{R}} (\omega, \omega') + Q_{\nu \mu}^{\mathrm{R}} (\omega, \omega - \omega') ) / 2!
, \\
Q_{\mu \nu}^{\mathrm{R}} (\omega, \omega')
	& = - \frac{1}{\hbar^2}  \int_{0}^{\infty} \mathrm{d}t \int_{0}^{\infty} \mathrm{d}t' e^{\zi (\omega + \zi \eta) t + \zi (\omega' + \zi \eta') t'}
		\braket{ \bigl[ [ \hat{B} (t + t'), \hat{A}_{\mu} (t') ], \hat{A}_{\nu} (0) \bigr] }
\label{eq:2nd_retarded_Green_function}
\end{align}
 We introduced the convergence factor $\eta$ and $\eta'$, but these two must have the relation $\eta > \eta'$ because of $Q_{\nu \mu}^{\mathrm{R}} (\omega, \omega - \omega')$.
 Hence, we use $2 \zi 0$ as the convergence factor for $\omega$ and $\zi 0$ as that for $\omega'$.

 As we show in Appendix~\ref{apx:sub:proof}, the corresponding correlation function in the Matsubara formulation to $\phi^{\mathrm{R}}_{\mu \nu} (\omega, \omega')$ (not to $Q_{\mu \nu}^{\mathrm{R}} (\omega, \omega')$) is given as
\begin{align}
\varphi_{\mu \nu} (\zi \omega_{\lambda}, \zi \omega_{\lambda'})
	& = \frac{1}{2! \hbar^2} \int_{0}^{\hbar \beta} \mathrm{d}\tau \int_{0}^{\hbar \beta} \mathrm{d}\tau'
		e^{\zi \omega_{\lambda} (\tau - \tau') + \zi \omega_{\lambda'} \tau'}
		\braket{ \mathrm{T}_{\tau,\tau'} \{ \hat{B} (\tau) \hat{A}_{\mu} (\tau') \hat{A}_{\nu} (0) \} }
\label{eq:second_order_response_Matsubara}
\end{align}
 Taking the analytic continuation $\zi \omega_{\lambda} \to \omega + 2 \zi 0$ and $\zi \omega_{\lambda'} \to \omega' + \zi 0$, the following relation is held
\begin{align}
\phi^{\mathrm{R}}_{\mu \nu} (\omega, \omega')
	& = \varphi_{\mu \nu} (\omega + 2 \zi 0, \omega' + \zi 0)
\label{eq:correspondence_2nd_order}
\end{align}
 Hence, we can evaluate the second order response [Eq.~(\ref{eq:second_order_response})] from the corresponding correlation function in the Matsubara formalism, $\varphi_{\mu \nu} (\zi \omega_{\lambda}, \zi \omega_{\lambda'})$, by taking the analytic continuations.

\subsection{\label{apx:sub:proof} Correspondence between $\varphi_{\mu \nu} (\zi \omega_{\lambda}, \zi \omega_{\lambda'})$ and $\phi^{\mathrm{R}}_{\mu \nu} (\omega, \omega')$}
 Here, we show Eq.~(\ref{eq:correspondence_2nd_order}).
 First, we perform the integrals of $t$ and $t'$ in $Q_{\mu \nu}^{\mathrm{R}} (\omega, \omega')$.
 Introducing $\omega_{+} = \omega + 2 \zi 0$ and $\omega'_{+} = \omega' + \zi 0$,
\begin{align}
Q_{\mu \nu}^{\mathrm{R}} (\omega, \omega')
	& = - \frac{1}{\hbar^2} \sum_{n,m,l} e^{\beta (\Omega - E_n)} ( 1 - e^{\beta ( E_n - E_l )} )
	\int_{0}^{\infty} \mathrm{d}t \int_{0}^{\infty} \mathrm{d}t' e^{\zi \omega_{+} t + \zi \omega'_{+} t'}
\notag \\ & \hspace{2em} \times
	\Bigl\{
		e^{\zi (E_n - E_m) t /\hbar + \zi (E_n - E_l) t' /\hbar }
			\bra{n} \hat{B} \ket{m} \bra{m} \hat{A}_{\mu} \ket{l} \bra{l} \hat{A}_{\nu} \ket{n}
\notag \\ & \hspace{3em}
		+ e^{- \zi (E_n - E_m) t /\hbar - \zi (E_n - E_l) t' /\hbar }
			\bra{m} \hat{B} \ket{n} \bra{n} \hat{A}_{\nu} \ket{l} \bra{l} \hat{A}_{\mu} \ket{m}
	\Bigr\}
\notag \\	& = \sum_{n,m,l} e^{\beta (\Omega - E_n)} ( 1 - e^{\beta ( E_n - E_l )} )
\notag \\ & \hspace{2em} \times
	\left\{
		\frac{ \bra{n} \hat{B} \ket{m} \bra{m} \hat{A}_{\mu} \ket{l} \bra{l} \hat{A}_{\nu} \ket{n} }{ ( \hbar \omega_{+} + E_n - E_m ) ( \hbar \omega'_{+} + E_n - E_l) }
		+ \frac{ \bra{m} \hat{B} \ket{n} \bra{n} \hat{A}_{\nu} \ket{l} \bra{l} \hat{A}_{\mu} \ket{m} }{ ( \hbar \omega_{+} - E_n + E_m ) ( \hbar \omega'_{+} - E_n + E_l) }
	\right\}
,\end{align}
and for $Q_{\nu \mu}^{\mathrm{R}} (\omega, \omega - \omega')$, by interchanging $n \leftrightarrow m$ in the above equation, we obtain
\begin{align*}
Q_{\nu \mu}^{\mathrm{R}} (\omega, \omega - \omega')
	& = \sum_{n,m,l} e^{\beta (\Omega - E_n)} e^{\beta (E_n - E_m)} ( 1 - e^{\beta ( E_m - E_l )} )
\\ & \hspace{-1em} \times
	\left\{
		\frac{ \bra{m} \hat{B} \ket{n} \bra{n} \hat{A}_{\nu} \ket{l} \bra{l} \hat{A}_{\mu} \ket{m} }{ ( \hbar \omega_{+} - E_n + E_m ) ( \hbar \omega_{+} - \hbar \omega'_{+} + E_m - E_l) }
		+ \frac{ \bra{n} \hat{B} \ket{m} \bra{m} \hat{A}_{\mu} \ket{l} \bra{l} \hat{A}_{\nu} \ket{n} }{ ( \hbar \omega_{+} + E_n - E_m ) ( \hbar \omega_{+} - \hbar \omega'_{+} - E_m + E_l) }
	\right\}
\end{align*}
 Here, $\omega_{+} - \omega'_{+} = \omega - \omega' + \zi (\eta - \eta')$, and $\eta > \eta'$ is needed for the convergence in the limit $t \to \infty$, hence putting $\eta = 2 \eta'$.
 From these, $\phi^{\mathrm{R}}_{\mu \nu} (\omega, \omega') = Q_{\mu \nu}^{\mathrm{R}} (\omega, \omega') + Q_{\nu \mu}^{\mathrm{R}} (\omega, \omega - \omega')$ is given as
\begin{align}
\phi^{\mathrm{R}}_{\mu \nu} (\omega, \omega')
	& = \frac{1}{2} \sum_{n,m,l} e^{\beta (\Omega - E_n)}
		\left(
			\frac{ \bra{n} \hat{B} \ket{m} \bra{m} \hat{A}_{\mu} \ket{l} \bra{l} \hat{A}_{\nu} \ket{n} }{ \hbar \omega_{+} + E_n - E_m }
				\mathrm{a} (\omega_{+}, \omega'_{+})
\right. \notag \\ & \left.
			+ \frac{ \bra{m} \hat{B} \ket{n} \bra{n} \hat{A}_{\nu} \ket{l} \bra{l} \hat{A}_{\mu} \ket{m} }{ - \hbar \omega_{+} + E_n - E_m }
				\mathrm{a} (- \omega_{+}, - \omega'_{+})
		\right)
\label{eq:phi^R_in_Lehmann_rep}
,\end{align}
where
\begin{align}
\mathrm{a} (\omega_{+}, \omega'_{+})
	& \equiv \frac{ 1 - e^{\beta (E_n - E_l)} }{ \hbar \omega'_{+} + E_n - E_l }
		+ \frac{ e^{\beta (E_n - E_m)} (1 - e^{\beta (E_m - E_l)}) }{ \hbar \omega_{+} - \hbar \omega'_{+} - E_m + E_l }
\end{align}

 Next, we perform the integrals of the correlation function in the Matsubara formalism [Eq.~(\ref{eq:second_order_response_Matsubara})].
 From the time-ordering operator, for $\tau > \tau' > 0$, 
\begin{align}
\braket{ \mathrm{T}_{\tau,\tau'} \{ \hat{B} (\tau) \hat{A}_{\mu} (\tau') \hat{A}_{\nu} (0) \} }
	& = \sum_{m,n,l} e^{\beta (\Omega - E_n)} e^{(E_m - E_l) \tau'/\hbar}
			\bra{n} \hat{B} \ket{m} \bra{m} \hat{A}_{\mu} \ket{l} \bra{l} \hat{A}_{\nu} \ket{n}
			e^{(E_n - E_m) \tau/\hbar}
,\end{align}
hence we find
\begin{align}
\frac{1}{\hbar^2}
& \int_{0}^{\hbar \beta} \mathrm{d}\tau'
	 \int_{\tau'}^{\hbar \beta} \mathrm{d}\tau
		e^{\zi \omega_{\lambda} (\tau - \tau') + \zi \omega_{\lambda'} \tau'}
	\braket{ \mathrm{T}_{\tau,\tau'} \{ \hat{B} (\tau) \hat{A}_{\mu} (\tau') \hat{A}_{\nu} (0) \} }
\notag \\
	& = \sum_{m, n, l} e^{\beta (\Omega - E_n)}
		\bra{n} \hat{B} \ket{m} \bra{m} \hat{A}_{\mu} \ket{l} \bra{l} \hat{A}_{\nu} \ket{n}
		\frac{1}{\hbar^2} \int_{0}^{\hbar \beta} \mathrm{d}\tau' e^{\zi (\omega_{\lambda'} - \omega_{\lambda}) \tau'} e^{(E_m - E_l) \tau'/\hbar}
			\int_{\tau'}^{\hbar \beta} \mathrm{d}\tau e^{(\zi \hbar \omega_{\lambda} + E_n - E_m) \tau / \hbar}
\notag \\
	& = \sum_{m, n, l} e^{\beta (\Omega - E_n)}
		\frac{ \bra{n} \hat{B} \ket{m} \bra{m} \hat{A}_{\mu} \ket{l} \bra{l} \hat{A}_{\nu} \ket{n} }{ \zi \hbar \omega_{\lambda} + E_n - E_m}
		\mathrm{a} (\zi \omega_{\lambda}, \zi \omega_{\lambda'})
\end{align}
 Also for $\tau' > \tau > 0$,
\begin{align}
\braket{ \mathrm{T}_{\tau,\tau'} \{ \hat{B} (\tau) \hat{A}_{\mu} (\tau') \hat{A}_{\nu} (0) \} }
	& = \sum_{m,n,l} e^{\beta (\Omega - E_l)} e^{(E_l - E_m) \tau'/\hbar}
			\bra{m} \hat{B} \ket{n} \bra{n} \hat{A}_{\nu} \ket{l} \bra{l} \hat{A}_{\mu} \ket{m}
			e^{(E_m - E_n) \tau/\hbar}
,\end{align}
and then, we obtain
\begin{align}
\frac{1}{\hbar^2}
& \int_{0}^{\hbar \beta} \mathrm{d}\tau'
	 \int_{0}^{\tau'} \mathrm{d}\tau
		e^{\zi \omega_{\lambda} (\tau - \tau') + \zi \omega_{\lambda'} \tau'}
	\braket{ \mathrm{T}_{\tau,\tau'} \{ \hat{B} (\tau) \hat{A}_{\mu} (\tau') \hat{A}_{\nu} (0) \} }
\notag \\
	& = \sum_{m,n,l} e^{\beta (\Omega - E_l)} 
			\bra{m} \hat{B} \ket{n} \bra{n} \hat{A}_{\nu} \ket{l} \bra{l} \hat{A}_{\mu} \ket{m}
		\frac{1}{\hbar^2} \int_{0}^{\hbar \beta} \mathrm{d}\tau' e^{\zi (\omega_{\lambda'} - \omega_{\lambda}) \tau'} e^{(E_l - E_m) \tau'/\hbar}
			\int_{0}^{\tau'} \mathrm{d}\tau e^{(\zi \hbar \omega_{\lambda} + E_m - E_n) \tau / \hbar}
\notag \\
	& = \sum_{m, n, l} e^{\beta (\Omega - E_n)}
		\frac{ \bra{m} \hat{B} \ket{n} \bra{n} \hat{A}_{\nu} \ket{l} \bra{l} \hat{A}_{\mu} \ket{m} }{ - \zi \hbar \omega_{\lambda} - E_m + E_n}
		\mathrm{a} (- \zi \omega_{\lambda}, - \zi \omega_{\lambda'})
\end{align}
 Therefore, Eq.~(\ref{eq:second_order_response_Matsubara}) is rewritten as
\begin{align}
\varphi_{\mu \nu} (\zi \omega_{\lambda}, \zi \omega_{\lambda'})
	& = \frac{1}{2} \sum_{n,m,l} e^{\beta (\Omega - E_n)}
		\left(
			\frac{ \bra{n} \hat{B} \ket{m} \bra{m} \hat{A}_{\mu} \ket{l} \bra{l} \hat{A}_{\nu} \ket{n} }{ \zi \hbar \omega_{\lambda} + E_n - E_m }
				\mathrm{a} (\zi \omega_{\lambda}, \zi \omega_{\lambda'})
\right. \notag \\ & \left.
			+ \frac{ \bra{m} \hat{B} \ket{n} \bra{n} \hat{A}_{\nu} \ket{l} \bra{l} \hat{A}_{\mu} \ket{m} }{ - \zi \hbar \omega_{\lambda} + E_n - E_m }
				\mathrm{a} (- \zi \omega_{\lambda}, - \zi \omega_{\lambda'})
		\right)
,\end{align}
and as compared with Eq.~(\ref{eq:phi^R_in_Lehmann_rep}), it is obvious that Eq.~(\ref{eq:correspondence_2nd_order}) is held.

\subsection{Third and higher order responses}
 The third order response, $k = 3$ for Eq.~(\ref{eq:apx:Kubo_formula}), reads
\begin{align}
\varDelta_3 B (t)
	& = \int_{-\infty}^{\infty} \mathrm{d}t_1 \int_{-\infty}^{\infty} \mathrm{d}t_2 \int_{-\infty}^{\infty} \mathrm{d}t_3
		\phi^{\rm R}_{\mu \nu \xi} (t, t_1, t_2, t_3)
		F_{\mu} (t_1) F_{\nu} (t_2) F_{\xi} (t_3)
,\end{align}
where $\phi^{\rm R}_{\mu \nu \xi} (t, t_1, t_2, t_3)$ is a symmetrized response coefficient given as
\begin{align}
\phi^{\rm R}_{\mu \nu \xi} (t, t_1, t_2, t_3)
	& = \frac{1}{3!} \Bigl(
		Q^{\mathrm{R}}_{\mu \nu \xi} (t, t_1, t_2, t_3)
		+ Q^{\mathrm{R}}_{\mu \xi \nu} (t, t_1, t_3, t_2)
		+ Q^{\mathrm{R}}_{\nu \mu \xi} (t, t_2, t_1, t_3)
\notag \\ & \hspace{2em}
		+ Q^{\mathrm{R}}_{\nu \xi \mu} (t, t_2, t_3, t_1)
		+ Q^{\mathrm{R}}_{\xi \nu \mu} (t, t_3, t_1, t_2)
		+ Q^{\mathrm{R}}_{\xi \mu \nu} (t, t_3, t_2, t_1)
	\Bigr)
, \\
Q^{\mathrm{R}}_{\mu \nu \xi} (t, t_1, t_2, t_3)
	& = \left( \frac{-1}{\zi \hbar} \right)^3
		\theta (t - t_1) \theta (t_1 - t_2) \theta (t_2 - t_3)
		\left\langle \Bigl[ \bigl[ \hat{B} (t), \hat{A}_{\mu} (t_1) ], \hat{A}_{\nu} (t_2) \bigr], \hat{A}_{\xi} (t_3) \Bigl] \right\rangle
.\end{align}
 One can see $Q^{\mathrm{R}}_{\mu \nu \xi} (t, t_1, t_2, t_3) = Q_{\mu \nu \xi}^{\mathrm{R}} (t - t_1, t_1 - t_2, t_2 - t_3)$ by using Eq.~(\ref{eq:expansion_by_eigenstates}), and the Fourier form is shown as
\begin{align}
Q^{\mathrm{R}}_{\mu \nu \xi} (\omega, \omega', \omega'')
	& = \left( \frac{-1}{\zi \hbar} \right)^3
		\iiint_{0}^{\infty} \mathrm{d}t \mathrm{d}t' \mathrm{d}t'' e^{\zi (\omega + \zi \eta) t + \zi (\omega' + \zi \eta') t' + \zi (\omega'' + \zi \eta'') t''}
\notag \\ & \hspace{4em} \times
		\left\langle \Bigl[ \bigl[ \hat{B} (t + t' + t''), \hat{A}_{\mu} (t' + t'') ], \hat{A}_{\nu} (t'') \bigr], \hat{A}_{\xi} (0) \Bigl] \right\rangle
.\end{align}
 Hence, the third order response in the Fourier space is given as
\begin{align}
\varDelta_3 B (\omega)
	& = \frac{1}{3!} \int_{-\infty}^{\infty} \frac{\mathrm{d}\omega'}{2 \pi} \int_{-\infty}^{\infty} \frac{\mathrm{d}\omega''}{2 \pi}
		\phi^{\rm R}_{\mu \nu \xi} (\omega, \omega', \omega'')
		F_{\mu} (\omega - \omega') F_{\nu} (\omega' - \omega'') F_{\xi} (\omega'')
\end{align}
with
\begin{align}
\phi^{\rm R}_{\mu \nu \xi} (\omega, \omega', \omega'')
	& = Q^{\mathrm{R}}_{\mu \nu \xi} (\omega, \omega', \omega'')
		+ Q^{\mathrm{R}}_{\mu \xi \nu} (\omega, \omega', \omega' - \omega'')
		+ Q^{\mathrm{R}}_{\nu \mu \xi} (\omega, \omega + \omega'' - \omega', \omega'')
\notag \\ & \hspace{-2em}
		+ Q^{\mathrm{R}}_{\nu \mu \xi} (\omega, \omega + \omega'' - \omega', \omega - \omega'')
		+ Q^{\mathrm{R}}_{\xi \mu \nu} (\omega, \omega - \omega'', \omega' - \omega'')
		+ Q^{\mathrm{R}}_{\xi \nu \mu} (\omega, \omega - \omega', \omega - \omega'')
\label{eq:symmetric_3rd_response_coeff_freq}
.\end{align}
 Equation~(\ref{eq:symmetric_3rd_response_coeff_freq}) leads that the convergence factors need to have the relation $\eta > \eta' > \eta''$.
 Hence, we put $\eta = 3 \eta''$, $\eta' =2 \eta''$.

 There is the corresponding correlation function in the Matsubara formalism to $\phi^{\rm R}_{\mu \nu \xi} (\omega, \omega', \omega'')$ given by
\begin{align}
\varphi_{\mu \nu \xi} (\zi \omega_{\lambda}, \zi \omega_{\lambda'}, \zi \omega_{\lambda''})
	& = \frac{1}{3! \hbar^3} \iiint_{0}^{\hbar \beta} \mathrm{d}\tau \mathrm{d}\tau' \mathrm{d}\tau''
		e^{ \zi \omega_{\lambda} (\tau - \tau') + \zi \omega_{\lambda'} (\tau' - \tau'') + \zi \omega_{\lambda''} \tau''}
	\braket{ \mathrm{T} \{ \hat{B} (\tau) \hat{A}_{\mu} (\tau') \hat{A}_{\nu} (\tau'') \hat{A}_{\xi} (0) \} }
\label{eq:imaginary_time_3_Green_func}
.\end{align}
 By taking the analytic continuations, $\zi \omega_{\lambda} \to \omega + 3 \zi 0$, $\zi \omega_{\lambda'} \to \omega' + 2 \zi 0$, $\zi \omega_{\lambda''} \to \omega'' + \zi 0$, we have
\begin{align}
\phi^{\rm R}_{\mu \nu \xi} (\omega, \omega', \omega'')
	& = \varphi_{\mu \nu \xi} (\omega + 3 \zi 0, \omega' + 2 \zi 0, \omega'' + \zi 0)
\label{eq:relation_Matsubara_to_real_3rd}
.\end{align}

 From the $k = 1, 2, 3$-th order responses, it is expected that the $k$-th order response is evaluated as follows:
 the response of $\hat{B}$ to the external forces is expressed as
\begin{align}
\varDelta_{k} B (t)
	& = \iint \cdots \int_{-\infty}^{\infty} \mathrm{d}t_1 \mathrm{d}t_2 \cdots \mathrm{d}t_k \,
		\phi^{\rm R}_{\mu_1 \mu_2 \cdots \mu_k} (t - t_1,  t_1 - t_2, \cdots, t_{k-1} - t_k)
		F_{\mu_1} (t_1) F_{\mu_2} (t_2) \cdots F_{\mu_k} (t_k)
,\end{align}
where $\phi^{\rm R}_{\mu_1 \mu_2 \cdots \mu_{k}} (t - t_1,  t_1 - t_2, \cdots, t_{k-1} - t_k)$ is the response coefficient already symmetrized, whose Fourier component $\phi^{\rm R}_{\mu_1 \mu_2 \cdots \mu_k} (\omega, \omega_1, \omega_2, \cdots, \omega_{k-1})$ is evaluated from the corresponding correlation function in the Matsubara formalism
\begin{align}
& \hspace{-1em}
\varphi_{\mu_1 \mu_2 \cdots \mu_k} (\zi \omega_{\lambda}, \zi \omega_{\lambda_1}, \zi \omega_{\lambda_2}, \cdots, \zi \omega_{\lambda_{k-1}})
\notag \\
& = \frac{1}{k! \hbar^k}
	\iint \cdots \iint_{0}^{\hbar \beta} \mathrm{d}\tau \mathrm{d}\tau_1 \cdots \mathrm{d}\tau_{k-2} \mathrm{d}\tau_{k-1}
	e^{\zi \omega_{\lambda} (\tau - \tau_1) + \zi \omega_{\lambda_1} (\tau_1 - \tau_2) + \cdots + \zi \omega_{\lambda_{k-2}} (\tau_{k-2} - \tau_{k-1}) + \zi \omega_{\lambda_{k-1}} \tau_{k-1}}
\notag \\ & \hspace{13em} \times
	\braket{ \mathrm{T} \{ \hat{B} (\tau) \hat{A}_{\mu_1} (\tau_1) \hat{A}_{\mu_2} (\tau_2) \cdots \hat{A}_{\mu_{k-2}} (\tau_{k-2}) \hat{A}_{\mu_{k-1}} (\tau_{k-1}) \hat{A}_{\mu_k} (0) \} }
\end{align}
by taking the analytic continuations, $\zi \omega_{\lambda} \to \omega + k \zi 0$, $\zi \omega_{\lambda_1} \to \omega_1 + (k-1) \zi 0$, $\cdots$, $\zi \omega_{\lambda_{k-2}} \to \omega_{k-2} + 2 \zi 0$, $\zi \omega_{\lambda_{k-1}} \to \omega_{k-1} + \zi 0$;
\begin{align}
\phi^{\rm R}_{\mu_1 \mu_2 \cdots \mu_k} (\omega, \omega_1, \omega_2, \cdots, \omega_{k-1})
	& = \varphi_{\mu_1 \mu_2 \cdots \mu_k} (\omega + k \zi 0, \omega_1 + (k-1)\zi 0, \omega_2 + (k-2) \zi 0, \cdots, \omega_{k-1} + \zi 0)
.\end{align}

\section{\label{apx:expressions}Expressions of diagrams}
\begin{figure}[hbtp]
	\centering
	\includegraphics[width=0.90\linewidth]{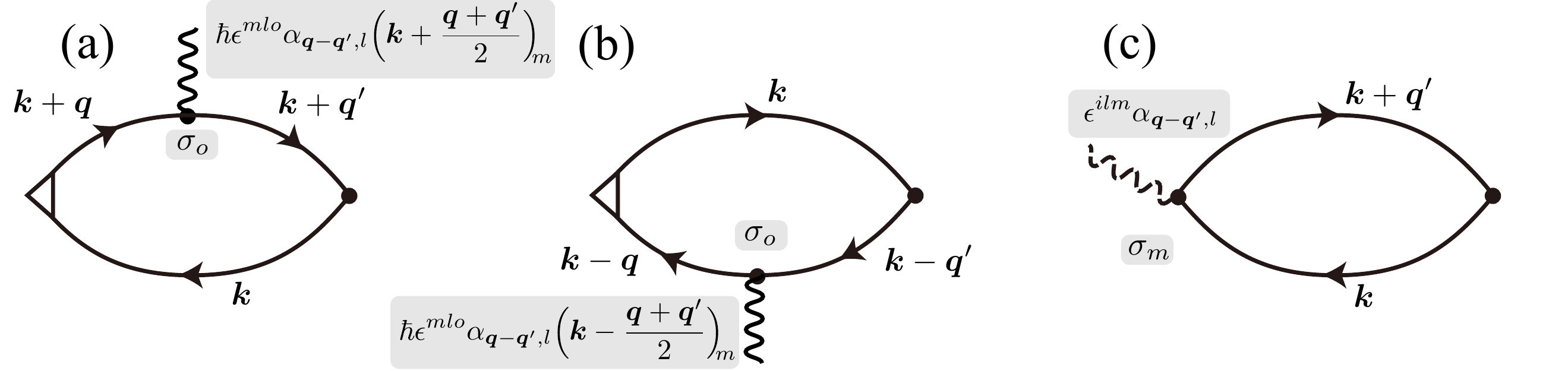}
	\includegraphics[width=0.45\linewidth]{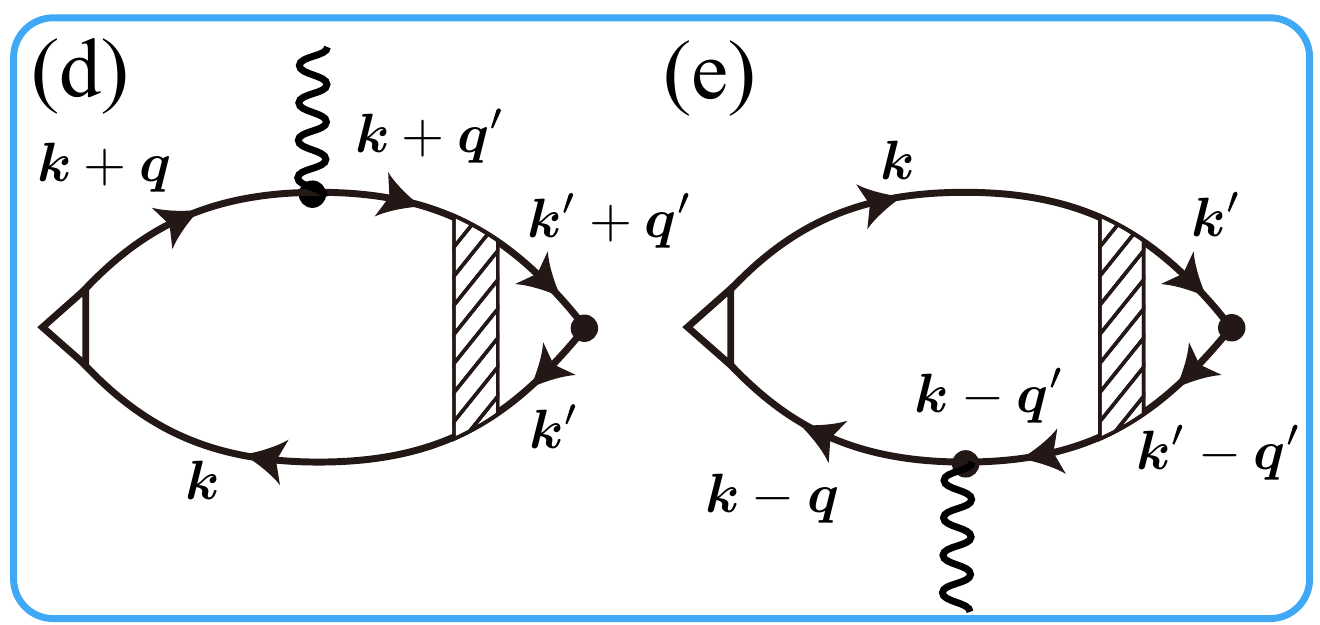}
	\includegraphics[width=0.45\linewidth]{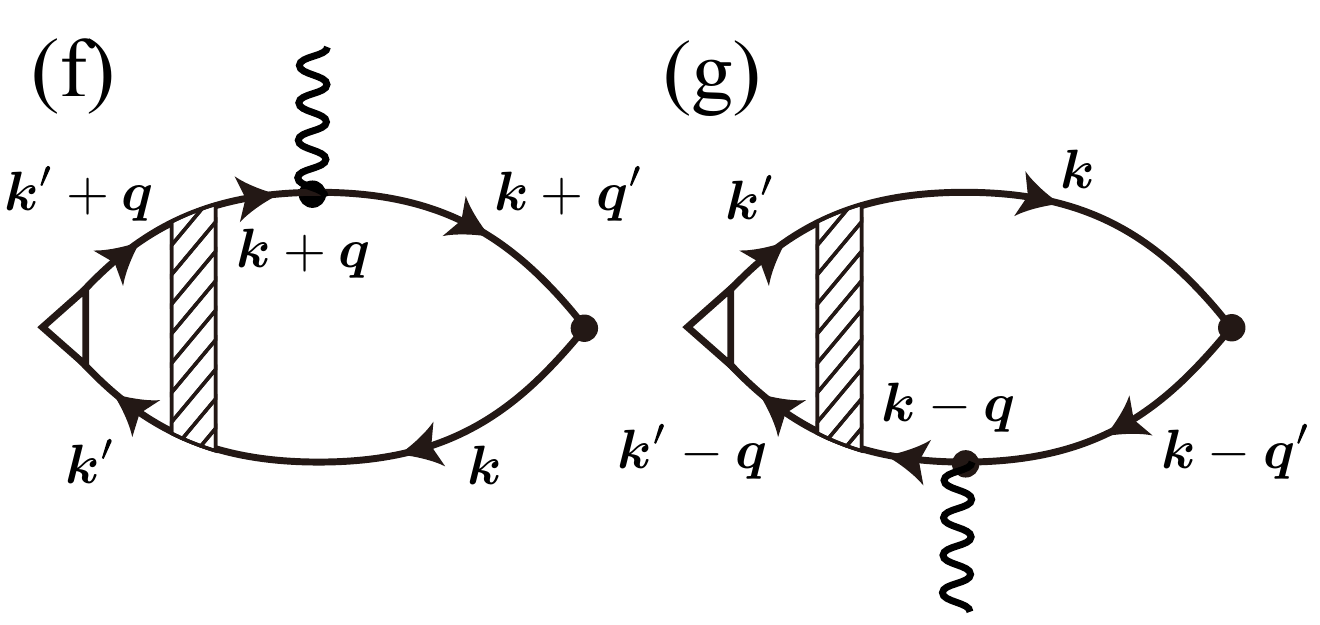}
	\includegraphics[width=0.45\linewidth]{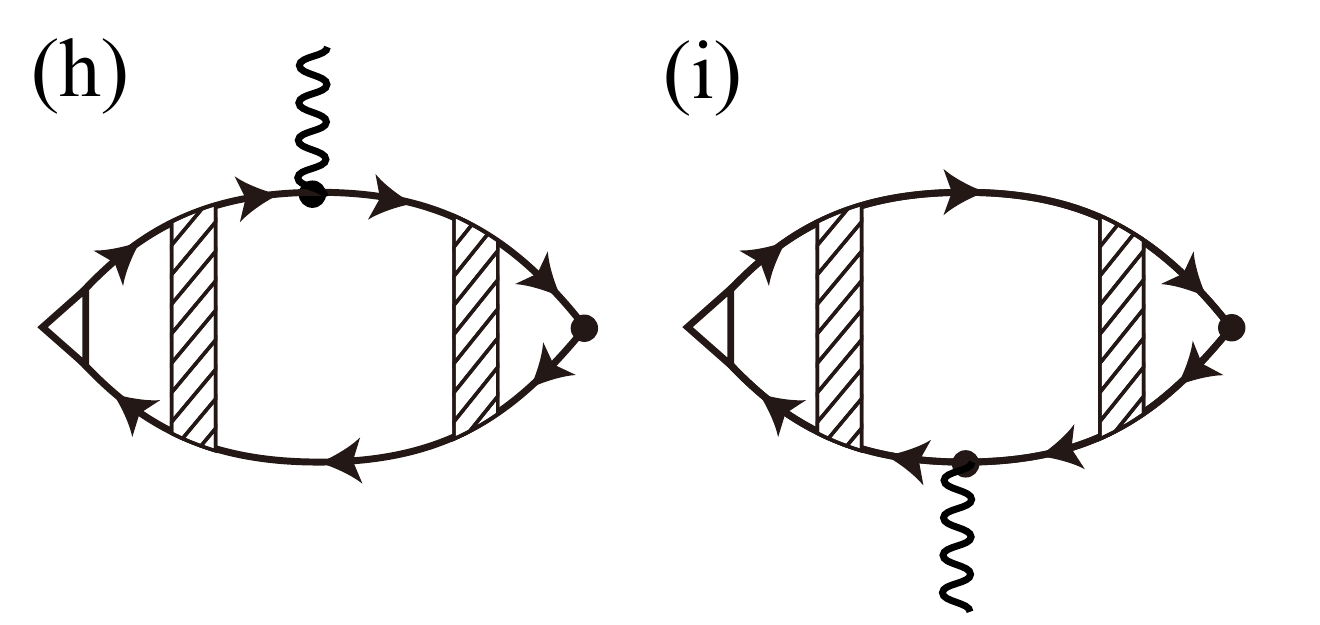}
	\includegraphics[width=0.45\linewidth]{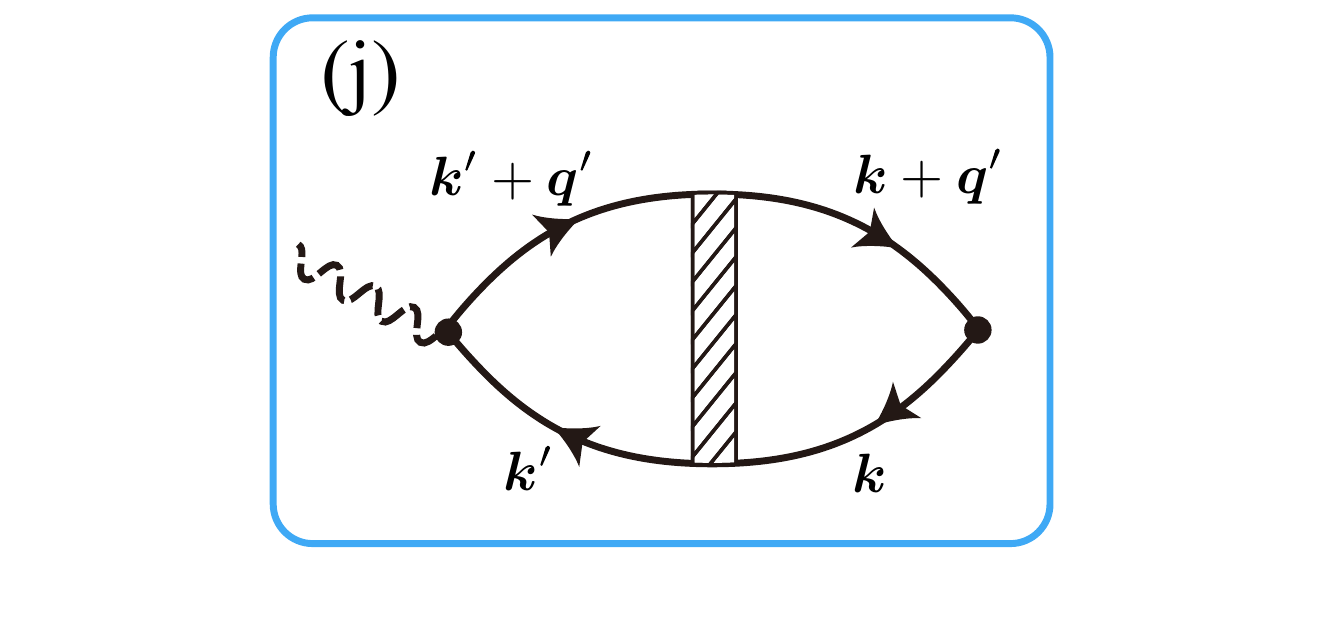}
	\caption{\label{fig:diagrams_chi_1}%
	The Feynman diagrams of $ \chi^{(1)}_{i j} (\zi \omega_{\lambda})$ in the first order of the Rashba spin-orbit interaction; (a)-(c) without the ladder type VCs and (d)-(j) with the VCs.
	The solid lines with arrows denote the Green functions without the Rashba interaction, given by Eq.~(\ref{eq:thermal_Green_function_H_0+H_R}), the filled circle represents the spin vertex, the unfilled triangle describes the normal velocity vertex, the dashed wavy line indicates the anomalous velocity vertex, and the solid wavy line depicts the Rashba-interaction vertex without the spin component.
	}
	\includegraphics[width=0.45\linewidth]{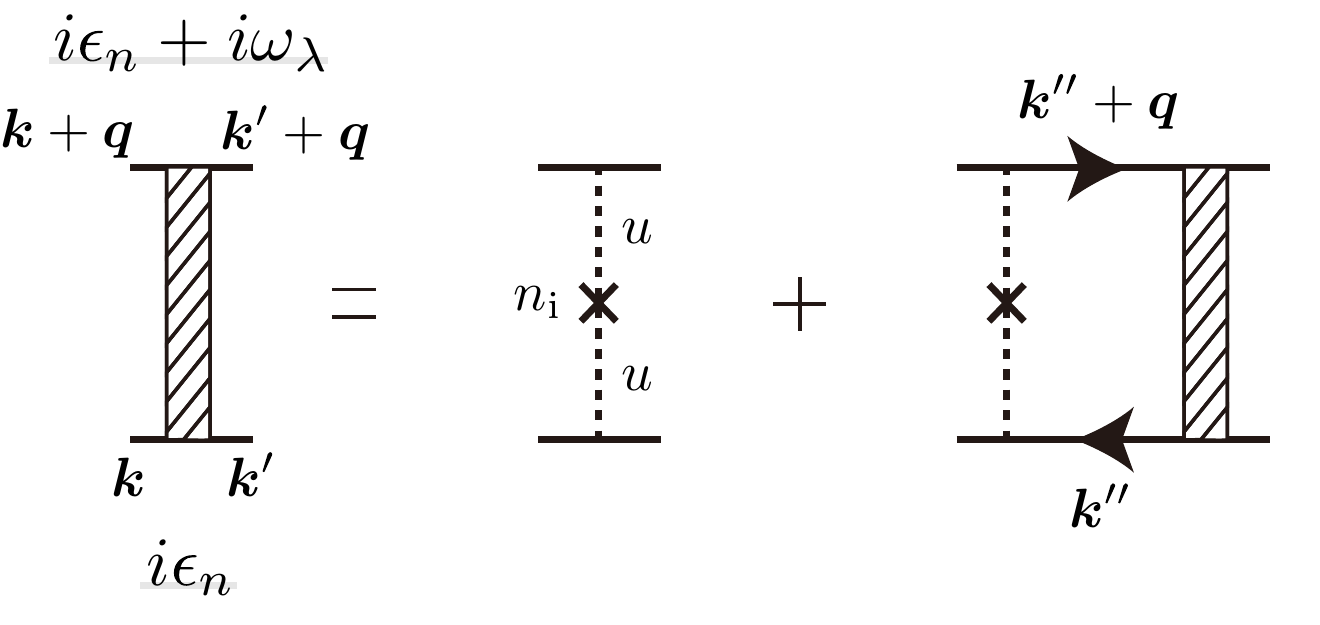}
	\caption{\label{fig:diffusion_ladder}%
	The diagrammatic description for the four-point vertex of the diffusion ladder.
	The dotted lines denote the impurity potential, and the cross symbol represents the impurity concentration.
	The solid lines without arrows are for the external momentums.
	}
\end{figure}
\begin{figure}[hbtp]
	\centering
	\includegraphics[width=\linewidth]{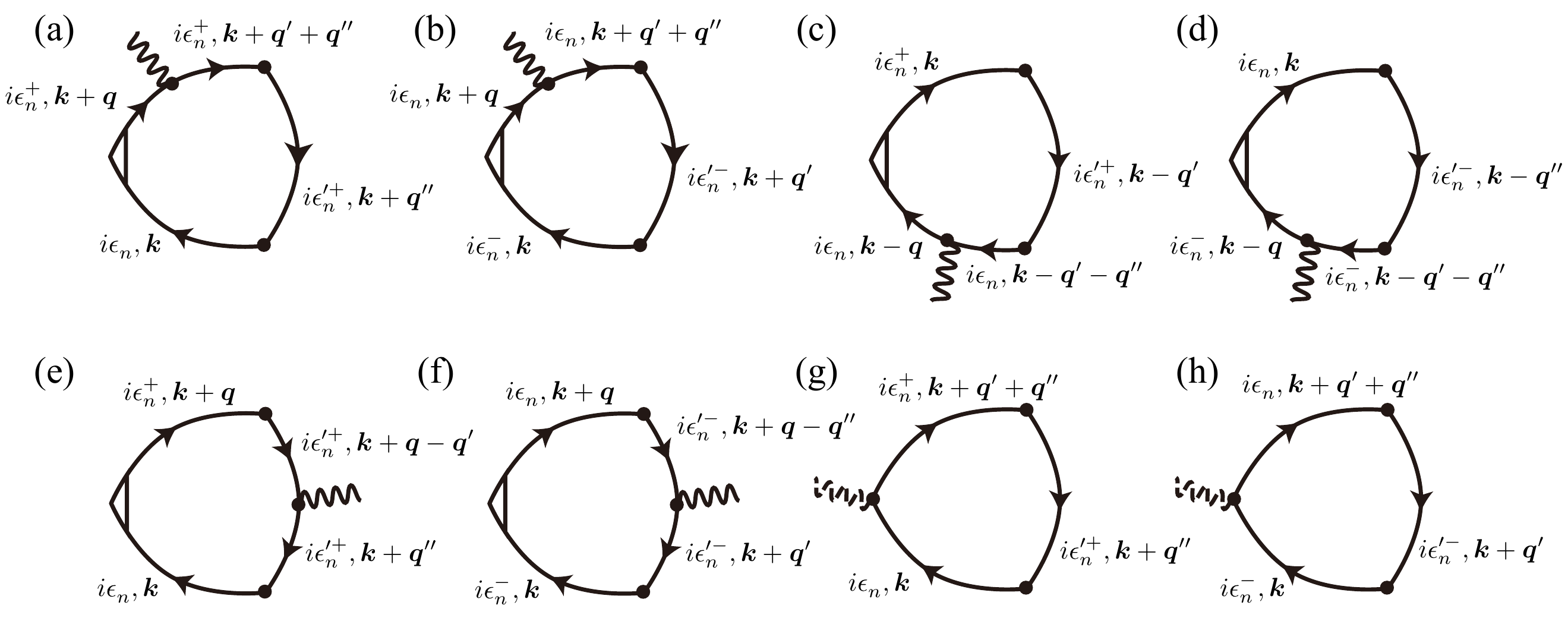}
	\caption{\label{fig:diagrams_chi_2}%
	The Feynman diagrams of $ \chi^{(2)}_{i j k} (\zi \omega_{\lambda}, \zi \omega_{\lambda'})$ in the first order of the Rashba interaction without the diffusion ladder VCs.
	The lines and symbols are defined in the caption of Fig.~\ref{fig:diagrams_chi_1}.
	(a)-(f): The contributions of the normal velocity term $\chi^{\rm (2,n)}_{i j k} (\zi \omega_{\lambda}, \zi \omega_{\lambda'})$ and (g)-(h): that of the anomalous velocity term $\chi^{\rm (2,a)}_{i j k} (\zi \omega_{\lambda}, \zi \omega_{\lambda'})$.
	Note that (b), (d), (f) and (h) are same contributions as that which are obtained by replacing $j \leftrightarrow k$, $\bq' \leftrightarrow \bq''$, $\epsilon_n^{} \to \epsilon_n^{-}$, $\epsilon_n^{+} \to \epsilon_n^{}$, and $\epsilon_n^{\prime +} \to \epsilon_n^{\prime -}$ in (a), (c), (e), and (g), respectively.
	}
\end{figure}

 In this Appendix, we show the expressions of all the diagrams contributing the non-local emergent electric fields shown in Fig.~\ref{fig:diagrams_chi_1} for the linear response and in Fig.~\ref{fig:diagrams_chi_2} for the second order response.

 Equations~(\ref{def:chi1n}) and (\ref{def:chi1a}) in the first order of the Rashba interaction are given respectively by Fig.~\ref{fig:diagrams_chi_1} (a)-(c), which reads
\begin{subequations}
\begin{align}
\chi^{\rm (1, n)}_{i j} (\bq, \bq';\zi \omega_{\lambda})
	& = e \epsilon^{m l j} \alpha_{\bq-\bq',l}
		\frac{1}{\beta} \sum_{n} \sum_{\sigma = \pm} \left(
			\delta_{j,z} \Pi^{i m, \sigma \sigma}_{\bq, \bq'} (\zi \epsilon_n^{+}, \zi \epsilon_n^{})
			+ \delta_{j,\perp} \Pi^{i m, \sigma \bar{\sigma}}_{\bq, \bq'} (\zi \epsilon_n^{+}, \zi \epsilon_n^{})
			+ (\zi \epsilon_n^{+} \leftrightarrow \zi \epsilon_n^{})
		\right)
\label{eq:chi_1n_expression}
, \\
\chi^{\rm (1, a)}_{i j} (\bq, \bq';\zi \omega_{\lambda})
	& = e \epsilon^{i l j} \alpha_{\bq-\bq',l}
		\frac{1}{\beta} \sum_{n} \sum_{\sigma = \pm} \left(
			\delta_{j,z} \varLambda^{\sigma \sigma}_{\bq'} (\zi \epsilon_n^{+}, \zi \epsilon_n^{})
			+ \delta_{j,\perp} \varLambda^{\sigma \bar{\sigma}}_{\bq'} (\zi \epsilon_n^{+}, \zi \epsilon_n^{})
		\right)
\label{eq:chi_1a_expression}
,\end{align}%
\end{subequations}
where $\zi \epsilon_n^{+} = \zi \epsilon_n + \zi \omega_{\lambda}$ and $\delta_{j, \perp} = (1 - \delta_{j,z})$, and
\begin{align}
\Pi^{i m, \sigma \sigma'}_{\bq,\bq'} (\zi \epsilon_m, \zi \epsilon_n)
	& = \frac{1}{V} \sum_{\bk} \frac{\hbar^2}{\me} \left( \bk + \frac{\bq}{2} \right)_i \left( \bk + \frac{\bq+\bq'}{2} \right)_m
			g_{\bk+\bq, \sigma} (\zi \epsilon_m)
			g_{\bk+\bq', \sigma'} (\zi \epsilon_m)
			g_{\bk, \sigma} (\zi \epsilon_n)
\label{eq:Pi^im}
, \\
\varLambda^{\sigma \sigma'}_{\bq'} (\zi \epsilon_m, \zi \epsilon_n)
	& = \frac{1}{V} \sum_{\bk}
		g_{\bk+\bq', \sigma} (\zi \epsilon_m)
		g_{\bk, \sigma'} (\zi \epsilon_n)
\label{eq:Lambda0}
\end{align}
 Here, we used $g_{-\bk, \sigma} (\zi \epsilon_n) = g_{\bk, \sigma} (\zi \epsilon_n)$ for calculating the diagram of Fig.\ref{fig:diagrams_chi_1}~(b), resulting in the term which is interchanged $\zi \epsilon_n^{+}$ and $\zi \epsilon_n^{}$ for the first two terms in Eq.~(\ref{eq:chi_1n_expression}).
 The Green function $g_{\bk, \sigma} (\zi \epsilon_n)$ is here expressed depending spin $\sigma = \pm$, but we will evaluate it as $g_{\bk, \sigma} (\zi \epsilon_n) = g_{\bk} (\zi \epsilon_n)$ in the next section.
 The four-point vertex of the diffusion ladder is given by Fig.~\ref{fig:diffusion_ladder},
\begin{align}
\varGamma^{\sigma \sigma'}_{\bq} (\zi \epsilon_n^{+}, \zi \epsilon_n^{})
	& = n_{\rm i} u^2
		+ \frac{(n_{\rm i} u^2)^2}{V} \sum_{\bk} g_{\bk+\bq, \sigma} (\zi \epsilon_n^{+}) g_{\bk, \sigma'} (\zi \epsilon_n)
		+ \frac{(n_{\rm i} u^2)^3}{V} \left( \sum_{\bk} g_{\bk+\bq, \sigma} (\zi \epsilon_n^{+}) g_{\bk, \sigma'} (\zi \epsilon_n) \right)^2
		+ \cdots
\notag \\
	& = \frac{ n_{\rm i} u^2 }{ 1 - n_{\rm i} u^2 \varLambda^{\sigma \sigma'}_{\bq} (\zi \epsilon_n^{+}, \zi \epsilon_n^{}) }
\label{eq:Gamma}
.\end{align}
 The diffusion ladder VCs of $\chi^{\rm (1, n)}_{i j}$ are given by Fig.~\ref{fig:diagrams_chi_1} (d)-(i), which read
\begin{align}
\chi^{\rm (1, n)(df)}_{i j} (\bq, \bq';\zi \omega_{\lambda})
	& = \chi^{\rm (d)+(e)}_{i j}
		+ \chi^{\rm (f)+(g)}_{i j}
		+ \chi^{\rm (h)+(i)}_{i j}
\label{eq:chi_1nVC_expression}
\end{align}
with
\begin{subequations}
\begin{align}
\chi^{\rm (d)+(e)}_{i j}
	& = e \epsilon^{m l j} \alpha_{\bq-\bq',l}
		\frac{1}{\beta} \sum_{n} \sum_{\sigma = \pm} \Bigl[
			\delta_{j,z} \Pi^{i m, \sigma \sigma}_{\bq, \bq'} (\zi \epsilon_n^{+}, \zi \epsilon_n^{})
			\varGamma^{\sigma \sigma}_{\bq'} (\zi \epsilon_n^{+}, \zi \epsilon_n^{})
			\varLambda^{\sigma \sigma}_{\bq'} (\zi \epsilon_n^{+}, \zi \epsilon_n^{})
\notag \\[-1ex] & \hspace{9em}
			+ \delta_{j,\perp} \Pi^{i m, \sigma \bar{\sigma}}_{\bq, \bq'} (\zi \epsilon_n^{+}, \zi \epsilon_n^{})
			\varGamma^{\bar{\sigma} \sigma}_{\bq'} (\zi \epsilon_n^{+}, \zi \epsilon_n^{})
			\varLambda^{\bar{\sigma} \sigma}_{\bq'} (\zi \epsilon_n^{+}, \zi \epsilon_n^{})
			+ (\zi \epsilon_n^{+} \leftrightarrow \zi \epsilon_n^{})
		\Bigr]
, \\
\chi^{\rm (f)+(g)}_{i j}
	& = e \epsilon^{m l j} \alpha_{\bq-\bq',l}
		\frac{1}{\beta} \sum_{n} \sum_{\sigma = \pm} \Bigl[
			\delta_{j,z} \varLambda^{i, \sigma}_{\bq} (\zi \epsilon_n^{+}, \zi \epsilon_n^{})
			\varGamma^{\sigma \sigma}_{\bq} (\zi \epsilon_n^{+}, \zi \epsilon_n^{})
			\Pi^{m, \sigma \sigma}_{\bq,\bq'} (\zi \epsilon_n^{+}, \zi \epsilon_n^{})
\notag \\[-1ex] & \hspace{9em}
			+ \delta_{j,\perp} \varLambda^{i, \sigma}_{\bq} (\zi \epsilon_n^{+}, \zi \epsilon_n^{})
			\varGamma^{\sigma \sigma}_{\bq} (\zi \epsilon_n^{+}, \zi \epsilon_n^{})
			\Pi^{m, \sigma \bar{\sigma}}_{\bq,\bq'} (\zi \epsilon_n^{+}, \zi \epsilon_n^{})
			+ (\zi \epsilon_n^{+} \leftrightarrow \zi \epsilon_n^{})
		\Bigr]
, \\
\chi^{\rm (h)+(i)}_{i j}
	& = e \epsilon^{m l j} \alpha_{\bq-\bq',l}
		\frac{1}{\beta} \sum_{n} \sum_{\sigma = \pm} \Bigl[
			\delta_{j,z} 
			\varLambda^{i, \sigma}_{\bq} (\zi \epsilon_n^{+}, \zi \epsilon_n^{})
			\varGamma^{\sigma \sigma}_{\bq} (\zi \epsilon_n^{+}, \zi \epsilon_n^{})
			\Pi^{m, \sigma \sigma}_{\bq,\bq'} (\zi \epsilon_n^{+}, \zi \epsilon_n^{})
			\varGamma^{\sigma \sigma}_{\bq'} (\zi \epsilon_n^{+}, \zi \epsilon_n^{})
			\varLambda^{\sigma \sigma}_{\bq'} (\zi \epsilon_n^{+}, \zi \epsilon_n^{})
\notag \\[-1ex] & \hspace{9em}
			+ \delta_{j,\perp} \varLambda^{i, \sigma}_{\bq} (\zi \epsilon_n^{+}, \zi \epsilon_n^{})
			\varGamma^{\sigma \sigma}_{\bq} (\zi \epsilon_n^{+}, \zi \epsilon_n^{})
			\Pi^{m, \sigma \bar{\sigma}}_{\bq,\bq'} (\zi \epsilon_n^{+}, \zi \epsilon_n^{})
			\varGamma^{\bar{\sigma} \sigma}_{\bq'} (\zi \epsilon_n^{+}, \zi \epsilon_n^{})
			\varLambda^{\bar{\sigma} \sigma}_{\bq'} (\zi \epsilon_n^{+}, \zi \epsilon_n^{})
\notag \\ & \hspace{9em}
			+ (\zi \epsilon_n^{+} \leftrightarrow \zi \epsilon_n^{})
		\Bigr]
,\end{align}%
\end{subequations}
where
\begin{align}
\Pi^{i, \sigma \sigma'}_{\bq,\bq'} (\zi \epsilon_m, \zi \epsilon_n)
	& = \frac{\hbar}{V} \sum_{\bk} \left( \bk + \frac{\bq+\bq'}{2} \right)_i
			g_{\bk+\bq, \sigma} (\zi \epsilon_m)
			g_{\bk+\bq', \sigma'} (\zi \epsilon_m)
			g_{\bk, \sigma} (\zi \epsilon_n)
, \\
\varLambda^{i, \sigma}_{\bq} (\zi \epsilon_m, \zi \epsilon_n)
	& = \frac{1}{V} \sum_{\bk} \frac{\hbar}{\me} \left( \bk + \frac{\bq}{2} \right)_i
			g_{\bk+\bq, \sigma} (\zi \epsilon_m)
			g_{\bk, \sigma} (\zi \epsilon_n)
,\end{align}
and the diffusion ladder VCs of $\chi^{\rm (1, a)}_{i j}$ is given by Fig.~\ref{fig:diagrams_chi_1} (j), which reads
\begin{align}
\chi^{\rm (1, a)(df)}_{i j} (\bq, \bq';\zi \omega_{\lambda})
	& = e \epsilon^{i l j} \alpha_{\bq-\bq',l}
		\frac{1}{\beta} \sum_{n} \sum_{\sigma = \pm} \Bigl[
			\delta_{j,z} \varLambda^{\sigma \sigma}_{\bq'} (\zi \epsilon_n^{+}, \zi \epsilon_n)
			\varGamma^{\sigma \sigma}_{\bq'} (\zi \epsilon_n^{+}, \zi \epsilon_n)
			\varLambda^{\sigma \sigma}_{\bq'} (\zi \epsilon_n^{+}, \zi \epsilon_n)
\notag \\[-1ex] & \hspace{9em}
			+ \delta_{j,\perp} \varLambda^{\sigma \bar{\sigma}}_{\bq'} (\zi \epsilon_n^{+}, \zi \epsilon_n)
			\varGamma^{\bar{\sigma} \sigma}_{\bq'} (\zi \epsilon_n^{+}, \zi \epsilon_n)
			\varLambda^{\bar{\sigma} \sigma}_{\bq'} (\zi \epsilon_n^{+}, \zi \epsilon_n)
		\Bigr]
\label{eq:chi_1aVC_expression}
.\end{align}
 We calculate all the above quantities in Appendix~\ref{apx:diffusion_ladder_VC_calc}.


\begin{figure}[hbtp]
	\centering
	\includegraphics[width=\linewidth]{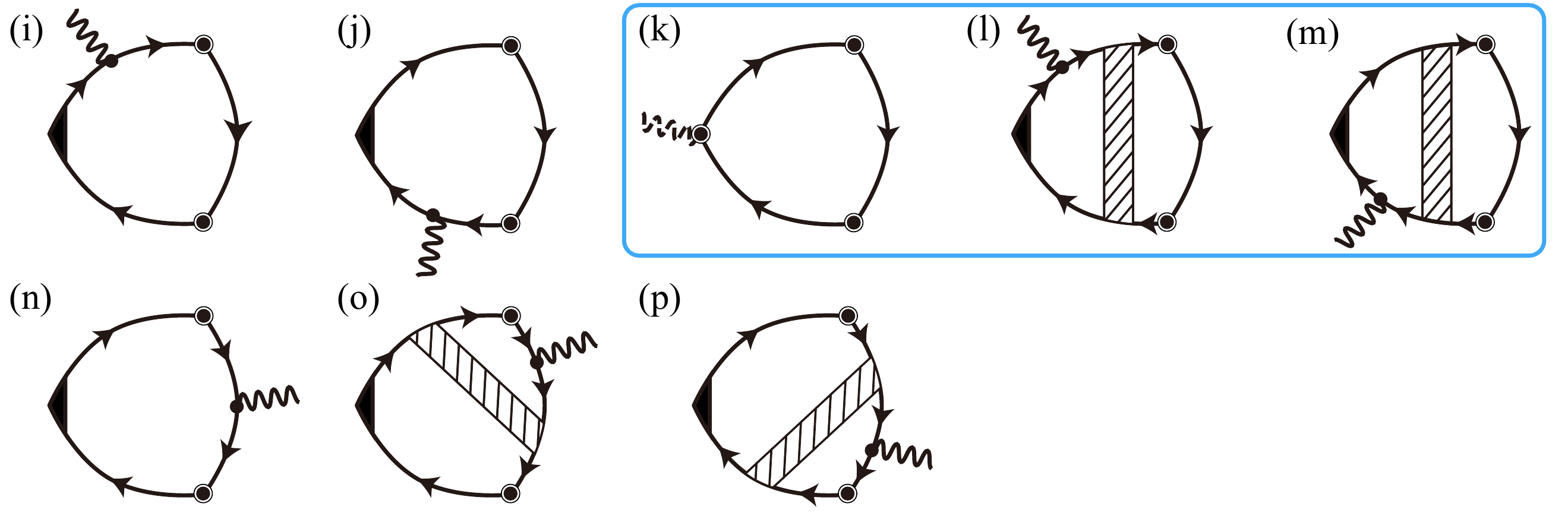}
	\includegraphics[width=\linewidth]{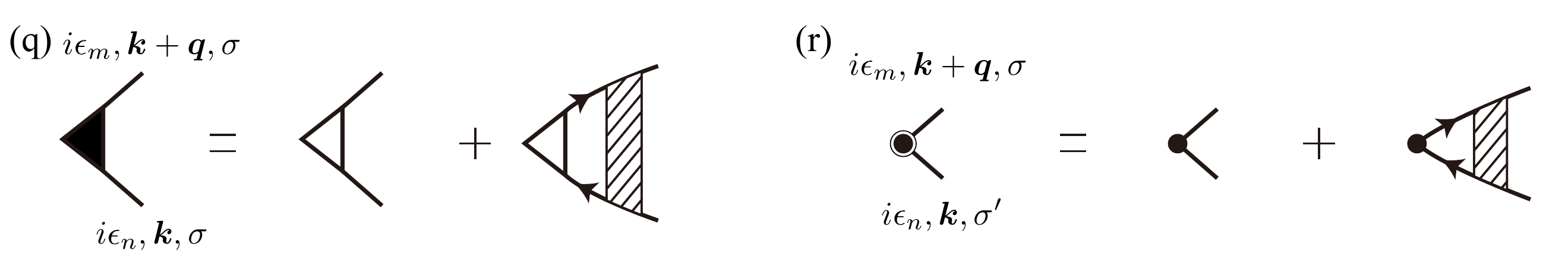}
	\caption{\label{fig:diagrams_chi_2_VCs}%
	All the Feynman diagrams of $ \chi^{(2)}_{i j k} (\zi \omega_{\lambda}, \zi \omega_{\lambda'})$ in the first order of the Rashba interaction;  (i), (j), (k) and (n) include the diagrams shown in Fig.~\ref{fig:diagrams_chi_2} (a)-(b), (c)-(d), (g)-(h) and (e)-(f), respectively.
	The three diagrams surrounded by a thick line include main contributions to the non-local emergent electric fields.
	In the diagrams (i)-(p), the momentums and Matsubara frequencies are not displayed for readability; they are same as in the diagrams of Fig.~\ref{fig:diagrams_chi_2} for (i), (j), (k), and (n).
	The momentums and Matsubara frequencies in (l), (m), (o) and (p) are expected from (i), (j) and (n) by using Fig.~\ref{fig:diffusion_ladder}.
	The filled triangle and double circle are the full vertexes of the normal velocity and the spin, respectively, given by (q) and (r).
	The momentums and Matsubara frequencies of the arrowed lines are shared with the corresponding diagrams in Fig.~\ref{fig:diagrams_chi_2}.
	}
\end{figure}
 For the second order response discussed in Sec.~\ref{sec:sub:second_order_response}, expanding Eqs.~(\ref{eq:chi_2n_def}) and (\ref{eq:chi_2a_def}) in the first order of the Rashba interaction, we have the diagrams in Fig.~\ref{fig:diagrams_chi_2}.
 The diagrams shown in Fig.~\ref{fig:diagrams_chi_2}~(a)-(f) and~(g)-(h) are obtained from Eqs.~(\ref{eq:chi_2n_def}) and (\ref{eq:chi_2a_def}), respectively, which reads
\begin{subequations}
\begin{align}
\chi^{\rm (2,n)}_{i j k} (\zi \omega_{\lambda}, \zi \omega_{\lambda'})
	& = \zi e \epsilon^{o l m} \epsilon^{m j k} \alpha_{\bq-\bq'-\bq'',l}
	\frac{1}{\beta} \sum_{n} \sum_{\sigma = \pm}
\notag \\[-1ex] & \hspace{1em} \times
	\Bigl[
		\delta_{m,z} \Bigl(
			\Xi^{i o, \sigma \sigma \bar{\sigma}}_{\bq,\bq',\bq''} (\zi \epsilon_n^{+}, \zi \epsilon_n^{\prime +}, \zi \epsilon_n^{})
			+ \Xi^{i o, \sigma \sigma \bar{\sigma}}_{\bq,\bq'',\bq'} (\zi \epsilon_n^{}, \zi \epsilon_n^{\prime +}, \zi \epsilon_n^{+})
			- \Theta^{i o, \sigma \bar{\sigma} \bar{\sigma}}_{\bq,\bq',\bq''} (\zi \epsilon_n^{+}, \zi \epsilon_n^{\prime +}, \zi \epsilon_n^{})
		\Bigr)
\notag \\ & \hspace{2em}
		+ \delta_{j,z} \Bigl(
			\Xi^{i o, \sigma \bar{\sigma} \bar{\sigma}}_{\bq,\bq',\bq''} (\zi \epsilon_n^{+}, \zi \epsilon_n^{\prime +}, \zi \epsilon_n^{})
			+ \Xi^{i o, \sigma \bar{\sigma} \sigma}_{\bq,\bq'',\bq'} (\zi \epsilon_n^{}, \zi \epsilon_n^{\prime +}, \zi \epsilon_n^{+})
			- \Theta^{i o, \sigma \sigma \bar{\sigma}}_{\bq,\bq',\bq''} (\zi \epsilon_n^{+}, \zi \epsilon_n^{\prime +}, \zi \epsilon_n^{})
		\Bigr)
\notag \\ & \hspace{2em}
		+ \delta_{k,z} \Bigl(
			\Xi^{i o, \sigma \bar{\sigma} \sigma}_{\bq,\bq',\bq''} (\zi \epsilon_n^{+}, \zi \epsilon_n^{\prime +}, \zi \epsilon_n^{})
			+ \Xi^{i o, \sigma \bar{\sigma} \bar{\sigma}}_{\bq,\bq'',\bq'} (\zi \epsilon_n^{}, \zi \epsilon_n^{\prime +}, \zi \epsilon_n^{+})
			- \Theta^{i o, \sigma \bar{\sigma} \sigma}_{\bq,\bq',\bq''} (\zi \epsilon_n^{+}, \zi \epsilon_n^{\prime +}, \zi \epsilon_n^{})
		\Bigr)
\notag \\ & \hspace{2em}
		+ (j \leftrightarrow k, \bq' \leftrightarrow \bq'', \epsilon_n^{} \to \epsilon_n^{-}, \epsilon_n^{+} \to \epsilon_n^{}, \epsilon_n^{\prime +} \to \epsilon_n^{\prime -})
	\Bigr]
\label{eq:chi_2n_expression}
, \\
\chi^{\rm (2,a)}_{i j k} (\zi \omega_{\lambda}, \zi \omega_{\lambda'})
	& = \zi e \epsilon^{i l m} \epsilon^{m j k} \alpha_{\bq-\bq'-\bq'',l}
	\frac{1}{\beta} \sum_{n} \sum_{\sigma = \pm}
	\Bigl[
		\delta_{m,z} \Lambda^{\sigma \bar{\sigma} \sigma}_{\bq',\bq''} (\zi \epsilon_n^{+}, \zi \epsilon_n^{\prime +}, \zi \epsilon_n^{})
		+ \delta_{j,z} \Lambda^{\sigma \sigma \bar{\sigma}}_{\bq',\bq''} (\zi \epsilon_n^{+}, \zi \epsilon_n^{\prime +}, \zi \epsilon_n^{})
\notag \\[-1ex] & \hspace{1em}
		+ \delta_{k,z} \Lambda^{\sigma \bar{\sigma} \bar{\sigma}}_{\bq',\bq''} (\zi \epsilon_n^{+}, \zi \epsilon_n^{\prime +}, \zi \epsilon_n^{})
		+ (j \leftrightarrow k, \bq' \leftrightarrow \bq'', \epsilon_n^{} \to \epsilon_n^{-}, \epsilon_n^{+} \to \epsilon_n^{}, \epsilon_n^{\prime +} \to \epsilon_n^{\prime -})
	\Bigr]
\label{eq:chi_2a_expression}
,\end{align}%
\end{subequations}
where $\zi \epsilon_n^{\pm} = \zi \epsilon_n \pm \zi \omega_{\lambda}$, $\zi \epsilon_n^{\prime \pm} = \zi \epsilon_n \pm \zi \omega_{\lambda'}$, and
\begin{subequations}
\begin{align}
\Xi^{i j, \sigma \sigma' \sigma''}_{\bp,\bq,\br} (\zi \epsilon_l, \zi \epsilon_m, \zi \epsilon_n)
	& = \frac{1}{V} \sum_{\bk} \frac{\hbar^2}{\me} \left( \bk + \frac{\bp}{2} \right)_i \left( \bk + \frac{\bp+\bq+\br}{2} \right)_j
\notag \\[-1ex] & \hspace{3em} \times
			g_{\bk+\bp, \sigma} (\zi \epsilon_l)
			g_{\bk+\bq+\br, \sigma'} (\zi \epsilon_l)
			g_{\bk+\br, \sigma''} (\zi \epsilon_m)
			g_{\bk, \sigma} (\zi \epsilon_n)
\label{eq:Xi}
, \\
\Theta^{i j, \sigma \sigma' \sigma''}_{\bp,\bq,\br} (\zi \epsilon_l, \zi \epsilon_m, \zi \epsilon_n)
	& = \frac{1}{V} \sum_{\bk} \frac{\hbar^2}{\me} \left( \bk + \frac{\bp}{2} \right)_i \left( \bk + \frac{\bp-\bq+\br}{2} \right)_j
\notag \\[-1ex] & \hspace{3em} \times
			g_{\bk+\bp, \sigma} (\zi \epsilon_l)
			g_{\bk+\bp-\bq, \sigma'} (\zi \epsilon_m)
			g_{\bk+\br, \sigma''} (\zi \epsilon_m)
			g_{\bk, \sigma} (\zi \epsilon_n)
\label{eq:Theta}
, \\
\Lambda^{\sigma \sigma' \sigma''}_{\bq,\br} (\zi \epsilon_l, \zi \epsilon_m, \zi \epsilon_n)
	& = \frac{1}{V} \sum_{\bk}
			g_{\bk+\bq+\br, \sigma} (\zi \epsilon_l)
			g_{\bk+\br, \sigma'} (\zi \epsilon_m)
			g_{\bk, \sigma''} (\zi \epsilon_n)
\label{eq:Lambda00}
\end{align}%
\end{subequations}
 The normal velocity terms containing diffusion ladder VCs are shown in Fig.~\ref{fig:diagrams_chi_2} (i)-(j), (l)-(p) and given as
\begin{align}
\chi^{\rm (2,n)(df)}_{i j k} (\zi \omega_{\lambda}, \zi \omega_{\lambda'})
	& = \chi^{\rm (i)+(j)}_{i j k}
		+ \chi^{\rm (l)+(m)}_{i j k}
		+ \chi^{\rm (n)+(o)+(p)}_{i j k}
\label{eq:chi_2nVC_expression}
, \\
\chi^{\rm (2,a)(df)}_{i j k} (\zi \omega_{\lambda}, \zi \omega_{\lambda'})
	& = \chi^{\rm (k)}_{i j k}
\label{eq:chi_2aVC_expression}
\end{align}
where 
\begin{align}
\chi^{\rm (l)+(m)}_{i j k}
	& = \zi e \epsilon^{o l m} \epsilon^{m j k} \alpha_{\bq-\bq'-\bq'',l}
	\frac{1}{\beta} \sum_{n} \sum_{\sigma = \pm}
\notag \\[-1ex] & \hspace{1em} \times
	\Bigl[
		\delta_{m,z} \Bigl(
			\Pi^{i o, \sigma \sigma}_{\bq,\bq'+\bq''} (\zi \epsilon_n^{+}, \zi \epsilon_n^{})
				\Gamma^{\sigma \sigma}_{\bq'+\bq''} (\zi \epsilon_n^{+}, \zi \epsilon_n^{})
				\Lambda^{\sigma \bar{\sigma} \sigma}_{\bq',\bq''} (\zi \epsilon_n^{+}, \zi \epsilon_n^{\prime +}, \zi \epsilon_n^{})
			+ (\zi \epsilon_n^{+} \leftrightarrow \zi \epsilon_n^{})
		\Bigr)
\notag \\ & \hspace{2em}
		+ \delta_{j,z} \Bigl(
			\Pi^{i o, \sigma \bar{\sigma}}_{\bq,\bq'+\bq''} (\zi \epsilon_n^{+}, \zi \epsilon_n^{})
				\Gamma^{\bar{\sigma} \sigma}_{\bq'+\bq''} (\zi \epsilon_n^{+}, \zi \epsilon_n^{})
				\Lambda^{\bar{\sigma} \bar{\sigma} \sigma}_{\bq',\bq''} (\zi \epsilon_n^{+}, \zi \epsilon_n^{\prime +}, \zi \epsilon_n^{})
\notag \\ & \hspace{4em}
			+ \Pi^{i o, \sigma \bar{\sigma}}_{\bq,\bq'+\bq''} (\zi \epsilon_n^{}, \zi \epsilon_n^{+})
				\Gamma^{\sigma \bar{\sigma}}_{\bq'+\bq''} (\zi \epsilon_n^{+}, \zi \epsilon_n^{})
				\Lambda^{\bar{\sigma} \sigma \sigma}_{\bq',\bq''} (\zi \epsilon_n^{}, \zi \epsilon_n^{\prime +}, \zi \epsilon_n^{+})
		\Bigr)
\notag \\ & \hspace{2em}
		+ \delta_{k,z} \Bigl(
			\Pi^{i o, \sigma \bar{\sigma}}_{\bq,\bq'+\bq''} (\zi \epsilon_n^{+}, \zi \epsilon_n^{})
				\Gamma^{\bar{\sigma} \sigma}_{\bq'+\bq''} (\zi \epsilon_n^{+}, \zi \epsilon_n^{})
				\Lambda^{\bar{\sigma} \sigma \sigma}_{\bq',\bq''} (\zi \epsilon_n^{+}, \zi \epsilon_n^{\prime +}, \zi \epsilon_n^{})
\notag \\ & \hspace{4em}
			+ \Pi^{i o, \sigma \bar{\sigma}}_{\bq,\bq'+\bq''} (\zi \epsilon_n^{}, \zi \epsilon_n^{+})
				\Gamma^{\sigma \bar{\sigma}}_{\bq'+\bq''} (\zi \epsilon_n^{+}, \zi \epsilon_n^{})
				\Lambda^{\bar{\sigma} \bar{\sigma} \sigma}_{\bq',\bq''} (\zi \epsilon_n^{}, \zi \epsilon_n^{\prime +}, \zi \epsilon_n^{+})
		\Bigr)
\notag \\ & \hspace{2em}
		+ (j \leftrightarrow k, \bq' \leftrightarrow \bq'', \epsilon_n^{} \to \epsilon_n^{-}, \epsilon_n^{+} \to \epsilon_n^{}, \epsilon_n^{\prime +} \to \epsilon_n^{\prime -})
	\Bigr]
\label{eq:chi-l+m}
, \\
\chi^{\rm (k)}_{i j k}
	& = \zi e \epsilon^{i l m} \epsilon^{m j k} \alpha_{\bq-\bq'-\bq'',l}
	\frac{1}{\beta} \sum_{n} \sum_{\sigma = \pm}
	\Bigl[
		\delta_{m,z}
			\Lambda^{\sigma \sigma}_{\bq'+\bq''} (\zi \epsilon_n^{+}, \zi \epsilon_n^{})
				\Gamma^{\sigma \sigma}_{\bq'+\bq''} (\zi \epsilon_n^{+}, \zi \epsilon_n^{})
				\Lambda^{\sigma \bar{\sigma} \sigma}_{\bq',\bq''} (\zi \epsilon_n^{+}, \zi \epsilon_n^{\prime +}, \zi \epsilon_n^{})
\notag \\[-1ex] & \hspace{12em}
		+ \delta_{j,z}
			\Lambda^{\sigma \bar{\sigma}}_{\bq'+\bq''} (\zi \epsilon_n^{+}, \zi \epsilon_n^{})
				\Gamma^{\sigma \bar{\sigma}}_{\bq'+\bq''} (\zi \epsilon_n^{+}, \zi \epsilon_n^{})
				\Lambda^{\sigma \sigma \bar{\sigma}}_{\bq',\bq''} (\zi \epsilon_n^{+}, \zi \epsilon_n^{\prime +}, \zi \epsilon_n^{})
\notag \\ & \hspace{12em}
		+ \delta_{k,z}
			\Lambda^{\sigma \bar{\sigma}}_{\bq'+\bq''} (\zi \epsilon_n^{+}, \zi \epsilon_n^{})
				\Gamma^{\sigma \bar{\sigma}}_{\bq'+\bq''} (\zi \epsilon_n^{+}, \zi \epsilon_n^{})
				\Lambda^{\sigma \bar{\sigma} \bar{\sigma}}_{\bq',\bq''} (\zi \epsilon_n^{+}, \zi \epsilon_n^{\prime +}, \zi \epsilon_n^{})
\notag \\ & \hspace{12em}
		+ (j \leftrightarrow k, \bq' \leftrightarrow \bq'', \epsilon_n^{} \to \epsilon_n^{-}, \epsilon_n^{+} \to \epsilon_n^{}, \epsilon_n^{\prime +} \to \epsilon_n^{\prime -})
	\Bigr]
\label{eq:chi-k}
,\end{align}
and $\chi^{\rm (i)+(j)}_{i j k}$ and $\chi^{\rm (n)+(o)+(p)}_{i j k}$ contains different types of diffusion from $\chi^{\rm (k)+(l)+(m)}_{i j k}$.

\section{\label{apx:diffusion_ladder_VC_calc}Calculation details}
 In this Appendix, we show the details of the calculations of the response coefficients at absolute zero, $T = 0$.
 In the present perturbative approach, the free Green functions are spin unpolarized, which means $g_{\bk, \sigma} (\zi \epsilon_n)$ is equivalent to $g_{\bk} (\zi \epsilon_n)$ defined as in Eq.~(\ref{eq:thermal_Green_function_H_0_delta-mu=0}).
 First, we calculate the liner response coefficient.
 Equations.~(\ref{eq:chi_1n_expression}) and (\ref{eq:chi_1a_expression}) are reduced to the following simple forms,
\begin{subequations}
\begin{align}
\chi^{\rm (1, n)}_{i j} (\bq, \bq';\zi \omega_{\lambda})
	& = 2 e \epsilon^{m l j} \alpha_{\bq-\bq',l}
		\frac{1}{\beta} \sum_{n} \left(
			\Pi^{i m}_{\bq, \bq'} (\zi \epsilon_n^{+}, \zi \epsilon_n^{})
			+ \Pi^{i m}_{\bq, \bq'} (\zi \epsilon_n^{}, \zi \epsilon_n^{+})
		\right)
\label{eq:chi_1n_expression_case1}
, \\
\chi^{\rm (1, a)}_{i j} (\bq, \bq';\zi \omega_{\lambda})
	& = 2 e \epsilon^{i l j} \alpha_{\bq-\bq',l}
		\frac{1}{\beta} \sum_{n}
			\varLambda^{}_{\bq'} (\zi \epsilon_n^{+}, \zi \epsilon_n^{})
\label{eq:chi_1a_expression_case1}
,\end{align}
and their diffusion VCs which mainly contribute to the non-local emergent electric fields are given as
\begin{align}
\chi^{\rm (1, n)(df)}_{i j} (\bq, \bq';\zi \omega_{\lambda})
	& \simeq \chi^{\rm (d)+(e)}_{i j}
\notag \\	
	& = 2 e \epsilon^{m l j} \alpha_{\bq-\bq',l}
		\frac{1}{\beta} \sum_{n} \Bigl[
			\Pi^{i m}_{\bq, \bq'} (\zi \epsilon_n^{+}, \zi \epsilon_n^{})
			\varGamma^{}_{\bq'} (\zi \epsilon_n^{+}, \zi \epsilon_n^{})
			\varLambda^{}_{\bq'} (\zi \epsilon_n^{+}, \zi \epsilon_n^{})
			+ (\zi \epsilon_n^{+} \leftrightarrow \zi \epsilon_n^{})
		\Bigr]
, \\
\chi^{\rm (1, a)(df)}_{i j} (\bq, \bq';\zi \omega_{\lambda})
	& = 2 e \epsilon^{i l j} \alpha_{\bq-\bq',l}
		\frac{1}{\beta} \sum_{n} \Bigl[
			\varLambda^{}_{\bq'} (\zi \epsilon_n^{+}, \zi \epsilon_n)
			\varGamma^{}_{\bq'} (\zi \epsilon_n^{+}, \zi \epsilon_n)
			\varLambda^{}_{\bq'} (\zi \epsilon_n^{+}, \zi \epsilon_n)
		\Bigr]
,\end{align}%
\end{subequations}
where $\Pi^{i m}_{\bq, \bq'} (\zi \epsilon_n^{+}, \zi \epsilon_n^{})$, $\varLambda^{}_{\bq'} (\zi \epsilon_n^{+}, \zi \epsilon_n)$, and $\varGamma^{}_{\bq'} (\zi \epsilon_n^{+}, \zi \epsilon_n^{})$ are given respectively by that dropped the spin dependences in Eqs.~(\ref{eq:Pi^im}), (\ref{eq:Lambda0}), and (\ref{eq:Gamma}).
 Using the following relations
\begin{align}
\varGamma^{}_{\bq'} (\zi \epsilon_n^{+}, \zi \epsilon_n^{})
\varLambda^{}_{\bq'} (\zi \epsilon_n^{+}, \zi \epsilon_n^{})
	& = - 1 + \frac{1}{1 - n_{\rm i} u^2 \varLambda^{}_{\bq'} (\zi \epsilon_n^{+}, \zi \epsilon_n^{})}
\label{eq:GammaLambda}
,\end{align}
and $\varLambda^{}_{\bq'} (\zi \epsilon_n^{+}, \zi \epsilon_n^{}) = \varLambda^{}_{\bq'} (\zi \epsilon_n^{}, \zi \epsilon_n^{+})$ due to $g_{-\bk} (\zi \epsilon_n^{}) = g_{\bk} (\zi \epsilon_n^{})$, we find
\begin{align}
\chi^{\rm (1)}_{i j} (\bq, \bq';\zi \omega_{\lambda})
	& \simeq 2 e \epsilon^{m l j} \alpha_{\bq-\bq',l}
		\frac{1}{\beta} \sum_{n}
			\frac{
				\Pi^{i m}_{\bq, \bq'} (\zi \epsilon_n^{+}, \zi \epsilon_n^{})
				+ \Pi^{i m}_{\bq, \bq'} (\zi \epsilon_n^{}, \zi \epsilon_n^{+})
				+ \delta_{i m} \varLambda^{}_{\bq'} (\zi \epsilon_n^{+}, \zi \epsilon_n)
			}{1 - n_{\rm i} u^2 \varLambda^{}_{\bq'} (\zi \epsilon_n^{+}, \zi \epsilon_n^{})}
\end{align}
 We rewrite the Matsubara summation of $\zi \epsilon_n$ to the contour integral, and then change the contour path as the two path of $[-\infty \pm \zi 0, +\infty \pm \zi 0]$ and $[-\infty - \zi \omega_{\lambda} \pm \zi 0, +\infty - \zi \omega_{\lambda} \pm \zi 0]$.
 After taking the analytic continuation $\zi \omega_{\lambda} \to \omega + \zi 0$, we obtain 
\begin{align}
\chi^{\mathrm{R}, \rm (1)}_{i j} (\bq, \bq'; \omega)
		& = 2 e \epsilon^{m l j} \alpha_{\bq-\bq',l} \left(
			\eta^{\rm (1)}_{\bq, \bq', i m}
			+ \zi \omega \, \varphi^{\rm (1)}_{\bq, \bq', i m}
			+ \cdots
		\right)
,\end{align}
where $\eta^{\rm (1)}_{\bq, \bq', i m}$ is the zeroth order term of $\omega$.
 The $\omega$-linear term $\varphi^{\rm (1)}_{\bq, \bq', i m}$ is obtained as
\begin{align}
\varphi^{\rm (1)}_{\bq, \bq', i m}
	& = \frac{\hbar}{2 \pi}
			\frac{
				2 \re \bigl[ \Pi^{i m}_{\bq, \bq'} (+\zi0,-\zi0) \bigr]
				+ \delta_{i m} \Lambda^{}_{\bq'} (+\zi0,-\zi0)
			}{1 - n_{\rm i} u^2 \Lambda^{}_{\bq'} (+\zi0,-\zi0)}
\label{eq:varphi^1}
, \\
\Pi^{i m}_{\bq, \bq'} (+\zi0,-\zi0)
	& = \frac{1}{V} \sum_{\bk} \frac{\hbar^2}{\me} \left( \bk + \frac{\bq}{2} \right)_i \left( \bk + \frac{\bq+\bq'}{2} \right)_m
			g^{\rm R}_{\bk+\bq}
			g^{\rm R}_{\bk+\bq'}
			g^{\rm A}_{\bk}
\label{eq:Pi^im_RA}
, \\
\Lambda^{}_{\bq'} (+\zi0,-\zi0)
	& = \frac{1}{V} \sum_{\bk}
		g^{\rm R}_{\bk+\bq'}
		g^{\rm A}_{\bk}
\label{eq:Lambda0_RA}
,\end{align}
where $g^{\rm R}_{\bk} = (-\epsilon_{\bk} + \mu + \zi \hbar/2\tau)^{-1}$ and $g^{\rm A}_{\bk} = (g^{\rm R}_{\bk})^{*}$.
 Here, we remained the terms which contains both the retarded and advanced Green functions in Eq.~(\ref{eq:varphi^1}).
 Expanding $\Pi^{i m}_{\bq, \bq'} (+\zi0,-\zi0)$ and $\Lambda^{}_{\bq'} (+\zi0,-\zi0)$ up to $q^2$ and $q'^2$ and performing the $\bk$-summations, we obtain
\begin{align}
\re \left[ \Pi^{i m}_{\bq, \bq'} (+\zi0,-\zi0) \right]
	& \simeq \delta_{i m} \left(
			- \frac{1}{2} I_{0 1 1}
			+ \frac{\hbar^2 (q^2 +q'^2)}{3 \me} \re\left[ I_{1 3 1} + \frac{4}{5} I_{2 4 1} \right]
			+ \frac{4 \hbar^2 \bq \cdot \bq'}{15 \me} \re \bigl[ I_{2 4 1} \bigr]
		\right)
\notag \\ & \hspace{1em}
		+ \frac{\hbar^2}{3 \me} (\bq + \bq')_i (\bq + \bq')_m \re\left[ I_{1 3 1} + \frac{4}{5} I_{2 4 1} \right]
		+ \frac{4 \hbar^2}{15 \me} (q_i q_m + q'_i q'_m) \re \bigl[  I_{2 4 1} \bigr]
\notag \\ & \hspace{1em}
		+ \frac{\hbar^2}{4 \me} q_i (q_m + q'_m) \re\left[ I_{0 2 1} + \frac{4}{3} I_{1 3 1} \right]
\\	& \simeq \frac{\pi \nu \tau}{\hbar} \left[
			\delta_{i m} \left( - 1 + D_0 \tau \bq \cdot \bq' \right)
			- D_0 \tau (q_i - q'_i) q'_m
		\right]
\label{eq:re_Pi^im_RA_result}
, \\
\Lambda^{}_{\bq'} (+\zi0,-\zi0)
	& \simeq I_{0 1 1}
		+ \frac{\hbar^2 q'^2}{2 \me} \left( I_{0 2 1} + \frac{4}{3} I_{1 3 1} \right)
\notag \\ &
	\simeq \frac{ 2 \pi \nu \tau}{\hbar} \left( 1 - D_0 q'^2 \tau \right)
\label{eq:Lambda0_RA_result}
,\end{align}
where $D_0 = 2 \eF \tau / 3 \me$ is the diffusion constant, and we used Eqs.~(\ref{eqs:I_lmn_evaluated}) and neglected the higher order contributions of $\hbar / \eF \tau \,(\ll 1)$.
 Hence, using Eq.~(\ref{eq:tau_def}), we find
\begin{align}
\varphi^{\rm (1)}_{\bq, \bq', i m}
	& = \frac{\nu \tau}{q'^2}
			\left( \delta_{i m} (\bq - \bq') \cdot \bq' - (q_i - q'_i) q'_m \right)
\label{eq:varphi1_im}
,\end{align}
which leads to Eq.~(\ref{eq:res:chi1}).
 It should be noted that $\varphi^{\rm (1)}_{\bq, \bq', i j m}$ does not contain any terms proportional to $1/D_0 q'^2$, which means that there is no contribution such as
\begin{align}
\braket{ \bm{j} (\br, t) }^{(1)}
	& \propto \int \frac{\bm{\alpha} (\br) \times \dot{\bm{M}} (\br', t)}{| \br - \br' |} \mathrm{d}\br'
.\end{align}

 Next, we calculate the second order response coefficient.
 The coefficient is also simplified in the case of $\mu_{+} = \mu_{-}$.
 Equations~(\ref{eq:chi_2n_expression}) and (\ref{eq:chi_2a_expression}) are given as
\begin{align}
\chi^{\rm (2,n)}_{i j k} (\zi \omega_{\lambda}, \zi \omega_{\lambda'})
	& = 2 \zi e \epsilon^{o l m} \epsilon^{m j k} \alpha_{\bq-\bq'-\bq'',l}
	\frac{1}{\beta} \sum_{n}
	\Bigl[
		\Xi^{i o}_{\bq,\bq',\bq''} (\zi \epsilon_n^{+}, \zi \epsilon_n^{\prime +}, \zi \epsilon_n^{})
		+ \Xi^{i o}_{\bq,\bq'',\bq'} (\zi \epsilon_n^{}, \zi \epsilon_n^{\prime +}, \zi \epsilon_n^{+})
\notag \\[-1ex] & \hspace{11em}
			- \Xi^{i o}_{\bq,\bq'',\bq'} (\zi \epsilon_n^{}, \zi \epsilon_n^{\prime -}, \zi \epsilon_n^{-})
			- \Xi^{i o}_{\bq,\bq',\bq''} (\zi \epsilon_n^{-}, \zi \epsilon_n^{\prime -}, \zi \epsilon_n^{})
\notag \\ & \hspace{11em}
			+ \Theta^{i o}_{\bq,\bq',\bq''} (\zi \epsilon_n^{+}, \zi \epsilon_n^{\prime +}, \zi \epsilon_n^{})
			- \Theta^{i o}_{\bq,\bq'',\bq'} (\zi \epsilon_n^{}, \zi \epsilon_n^{\prime -}, \zi \epsilon_n^{-})
	\Bigr]
\label{eq:chi_2n_expression_case1}
, \\
\chi^{\rm (2,a)}_{i j k} (\zi \omega_{\lambda}, \zi \omega_{\lambda'})
	& = 2 \zi e \epsilon^{i l m} \epsilon^{m j k} \alpha_{\bq-\bq'-\bq'',l}
	\frac{1}{\beta} \sum_{n}
	\Bigl[
		\Lambda^{}_{\bq',\bq''} (\zi \epsilon_n^{+}, \zi \epsilon_n^{\prime +}, \zi \epsilon_n^{})
		- \Lambda^{}_{\bq'',\bq'} (\zi \epsilon_n^{}, \zi \epsilon_n^{\prime -}, \zi \epsilon_n^{-})
	\Bigr]
\label{eq:chi_2a_expression_case1}
,\end{align}
where $\Xi^{i o}_{\bq,\bq',\bq''} (\zi \epsilon_n^{+}, \zi \epsilon_n^{\prime +}, \zi \epsilon_n^{})$, $\Theta^{i o}_{\bq,\bq',\bq''} (\zi \epsilon_n^{+}, \zi \epsilon_n^{\prime +}, \zi \epsilon_n^{})$, and $\Lambda^{}_{\bq',\bq''} (\zi \epsilon_n^{+}, \zi \epsilon_n^{\prime +}, \zi \epsilon_n^{})$ are same respectively as that dropped the spin indexes in Eqs.~(\ref{eq:Xi}), (\ref{eq:Theta}), and (\ref{eq:Lambda00}).
 However, Eqs.~(\ref{eq:chi_2n_expression_case1}) does not contribute the non-local emergent electric fields of our interest because it is canceled by the diffusion ladder VCs shown in Fig.~\ref{fig:diagrams_chi_2}~(i), (j), and (n).
 We also find that Eq.~(\ref{eq:chi_2a_expression_case1}) is canceled by the diffusion ladder VCs of the spin vertex $\sigma^j$ or $\sigma^k$, which gives rise to $1/D_0 q''^2$ or $1/D_0 q'^2$ and do not contribute the emergent electric field we focus in this paper.

 The main contributions of the diffusion VCs [Eqs.~(\ref{eq:chi_2nVC_expression}) and (\ref{eq:chi_2aVC_expression})] to the non-local emergent electric field are given as
\begin{align}
\chi^{\rm (2,n)(df)}_{i j k} (\zi \omega_{\lambda}, \zi \omega_{\lambda'})
	& = 2 \zi e \epsilon^{o l m} \epsilon^{m j k} \alpha_{\bq-\bq'-\bq'',l}
	\frac{1}{\beta} \sum_{n}
	\Bigl[
		\Pi^{i o}_{\bq,\bq'+\bq''} (\zi \epsilon_n^{+}, \zi \epsilon_n^{})
			\Gamma^{}_{\bq'+\bq''} (\zi \epsilon_n^{+}, \zi \epsilon_n^{})
			\Lambda^{}_{\bq',\bq''} (\zi \epsilon_n^{+}, \zi \epsilon_n^{\prime +}, \zi \epsilon_n^{})
\notag \\[-1ex] & \hspace{11em}
		- \Pi^{i o}_{\bq,\bq'+\bq''} (\zi \epsilon_n^{}, \zi \epsilon_n^{-})
			\Gamma^{}_{\bq'+\bq''} (\zi \epsilon_n^{}, \zi \epsilon_n^{-})
			\Lambda^{}_{\bq'',\bq'} (\zi \epsilon_n^{}, \zi \epsilon_n^{\prime -}, \zi \epsilon_n^{-})
\notag \\ & \hspace{11em}
		+ \Pi^{i o}_{\bq,\bq'+\bq''} (\zi \epsilon_n^{}, \zi \epsilon_n^{+})
			\Gamma^{}_{\bq'+\bq''} (\zi \epsilon_n^{}, \zi \epsilon_n^{+})
			\Lambda^{}_{\bq',\bq''} (\zi \epsilon_n^{}, \zi \epsilon_n^{\prime +}, \zi \epsilon_n^{+})
\notag \\ & \hspace{11em}
		- \Pi^{i o}_{\bq,\bq'+\bq''} (\zi \epsilon_n^{-}, \zi \epsilon_n^{})
			\Gamma^{}_{\bq'+\bq''} (\zi \epsilon_n^{-}, \zi \epsilon_n^{})
			\Lambda^{}_{\bq'',\bq'} (\zi \epsilon_n^{-}, \zi \epsilon_n^{\prime -}, \zi \epsilon_n^{})
	\Bigr]
\label{eq:chi_2nVC_expression_case1}
, \\
\chi^{\rm (2,a)(df)}_{i j k} (\zi \omega_{\lambda}, \zi \omega_{\lambda'})
	& = 2 \zi e \epsilon^{i l m} \epsilon^{m j k} \alpha_{\bq-\bq'-\bq'',l}
	\frac{1}{\beta} \sum_{n}
	\Bigl[
		\Lambda^{}_{\bq'+\bq''} (\zi \epsilon_n^{+}, \zi \epsilon_n^{})
			\Gamma^{}_{\bq'+\bq''} (\zi \epsilon_n^{+}, \zi \epsilon_n^{})
			\Lambda^{}_{\bq',\bq''} (\zi \epsilon_n^{+}, \zi \epsilon_n^{\prime +}, \zi \epsilon_n^{})
\notag \\[-1ex] & \hspace{11em}
		- \Lambda^{}_{\bq'+\bq''} (\zi \epsilon_n^{}, \zi \epsilon_n^{-})
			\Gamma^{}_{\bq'+\bq''} (\zi \epsilon_n^{}, \zi \epsilon_n^{-})
			\Lambda^{}_{\bq'',\bq'} (\zi \epsilon_n^{}, \zi \epsilon_n^{\prime -}, \zi \epsilon_n^{-})
	\Bigr]
\label{eq:chi_2aVC_expression_case1}
,\end{align}
where $\Lambda^{}_{\bq} (\zi \epsilon_n^{+}, \zi \epsilon_n^{})$ and $\Gamma^{}_{\bq} (\zi \epsilon_n^{+}, \zi \epsilon_n^{})$ are defined respectively by that dropped the spin indexes in Eqs.~(\ref{eq:Lambda0}) and (\ref{eq:Gamma}).
 For Eqs.~(\ref{eq:chi_2nVC_expression_case1}) and (\ref{eq:chi_2aVC_expression_case1}), we perform the similar procedures as in the calculations of the linear response coefficient; rewriting the Matsubara summation of $\zi \epsilon_n$ to the contour integral, changing the integral path into the three paths $[-\infty - \zi P_i \pm \zi 0, + \infty - \zi P_i \pm \zi 0]$ with $P_1 = \omega_{\lambda}$, $P_2 = \omega_{\lambda'}$ and $P_3 = 0$ for the terms which depend on the frequencies set $(\zi \epsilon_n^{+}, \zi \epsilon_n^{\prime +}, \zi \epsilon_n^{})$, and with $P_1 = 0$, $P_2 = - \omega_{\lambda'}$ and $P_3 = -\omega_{\lambda}$ for the terms which depend on the frequencies set $(\zi \epsilon_n^{}, \zi \epsilon_n^{\prime -}, \zi \epsilon_n^{-})$.
 Then, taking the analytic continuations as Eq.~(\ref{eq:analytic_continuation}), we obtain
\begin{align}
\chi^{\rm R, (2)}_{i j k} (\bq, \bq', \bq''; \omega, \omega')
	& = 2 \zi e \epsilon^{o l m} \epsilon^{m j k} \alpha_{\bq-\bq'-\bq'',l}
		\left(
		\eta^{(2)}_{\bq, \bq', \bq'', i o}
		+ \zi \omega \, \vartheta^{(2)}_{\bq, \bq', \bq'', i o}
		+ \zi \omega' \, \varphi^{(2)}_{\bq, \bq', \bq'', i o}
		+ \cdots
		\right)
\label{eq:chi^R2_expansion}
,\end{align}
where the first term gives rise to the current depending on the magnetization $\sim M_j (t) M_k (t)$, not on its dynamics, and the second term leads to the current due to the total derivative of the magnetizations $\sim \mathrm{d} (M_j (t) M_k (t)) / \mathrm{d}t$.
 The third term of Eq.~(\ref{eq:chi^R2_expansion}) is the component that we are focusing, which is obtained as
\begin{align}
\varphi^{(2)}_{\bq, \bq', \bq'', i o}
	& = \varphi^{(1)}_{\bq, \bq'+\bq'', i o}
		\frac{n_{\rm i} u^2}{2} \Phi_{\bq', \bq''}
,\end{align}
\begin{align}
\Phi_{\bq', \bq''}
	& = \sum_{s, t = \pm} s \Lambda_{\bq',\bq''} (\zi t0,\zi s0,-\zi t0)
	= \frac{2 \zi}{V} \sum_{\bk} \im \left[ ( g^{\rm R}_{\bk} - g^{\rm A}_{\bk} ) g^{\rm R}_{\bk+\bq'} g^{\rm A}_{\bk-\bq''} \right]
\end{align}
where $\varphi^{(1)}_{\bq, \bq'+\bq'', i o}$ is given by Eq.~(\ref{eq:varphi^1}), we neglected the terms which contains only retarded/advanced Green functions because they are just higher order contributions with respect to $\hbar/\eF \tau$, and we used $\Lambda_{\bq'',\bq'} (-\zi0,\pm\zi0,\zi0) = \Lambda_{\bq',\bq''} (\zi0,\pm\zi0,-\zi0)$.
 Here, expanding $\Phi_{\bq', \bq''}$ with respect to $\bq'$ and $\bq''$ up to the second order, we have
\begin{align}
\Phi_{\bq', \bq''}
	& = - 2 \zi \, \im\! \left[
		2 I_{0 1 2}
		+ \frac{\hbar^2 ( 3 q'^2 + 4 \bq' \cdot \bq'' + 3 q''^2)}{2 \me} I_{0 1 3}
		+ \frac{4 \hbar^2 (2 q'^2 + 3 \bq' \cdot \bq'' + 2 q''^2)}{3 \me} I_{1 1 4}
	\right]
\notag \\
	& \simeq - \frac{8 \zi \pi \nu \tau^2}{\hbar^2} \left\{
		1
		- D_0 \tau (2 q'^2 + 3 \bq' \cdot \bq'' + 2 q''^2)
	\right\}
,\end{align}
where $I_{l m n}$ is given by Eq.~(\ref{eq:I_lmn}) and left in the leading order of $\eF \tau / \hbar$ in the second equal by means of Eqs.~(\ref{eqs:I_lmn_evaluated}).
 Hence, using Eqs.~(\ref{eq:re_Pi^im_RA_result}), (\ref{eq:Lambda0_RA_result}), and (\ref{eq:tau_def}), we finally find
\begin{align}
\varphi^{(2)}_{\bq, \bq', \bq'', i o}
	& = - \frac{2 \zi \nu \tau^2}{\hbar} \frac{\delta_{i o} (\bq - \bq' - \bq'') \cdot (\bq' + \bq'') - (q_i - q'_i - q''_i) (q'_o + q''_o) }{(\bq' + \bq'')^2}
	+ \mathcal{O} (q^2, q'^2, q''^2)
,\end{align}
which leads to Eq.~(\ref{eq:res:chi2}).

\section{\label{apx:Integrals}Integrals}
 In this Appendix, we show  $\bk$-integrals $I_{l m n}$ that we use in this paper;
\begin{align}
I_{l m n}
	& = \frac{1}{V} \sum_{\bk} \left(\frac{\hbar^2 k^2}{2 \me}\right)^l \bigl( g^{\rm R}_{\bk} \bigr)^m \bigl( g^{\rm A}_{\bk} \bigr)^n
\label{eq:I_lmn}
,\end{align}
where we set $m + n \ge l + 3/2$ to its convergence.
 We rewrite the summation over $\bk$ into the energy integral,
\begin{align}
I_{l m n}
	& = \int_{-\eF}^{\infty} \mathrm{d}\xi
		\frac{ \nu (\xi + \eF) (\xi + \eF)^l }{(- \xi + \zi \hbar / 2 \tau)^m (- \xi - \zi \hbar / 2 \tau)^n}
\label{eq:I^A}
,\end{align}
where $\nu (\epsilon) \propto \sqrt{\epsilon}$ is DOS.
 For evaluating $I_{l m n}$ with respect to the leading order of $\hbar / \eF \tau$, it is valid to approximate $\xi + \eF \simeq \eF$, which means that DOS and the energy approximates to the values at the Fermi level, and to regard the lower limit of the integral as $-\infty$.
 However, we need to evaluate the higher order contributions precisely such as in Eq.~(\ref{eq:res:chi1}).
 Hence, we calculate $I_{l m n}$ without any approximations.

\begin{figure}[htbp!]
\centering
\includegraphics[width=0.4\linewidth]{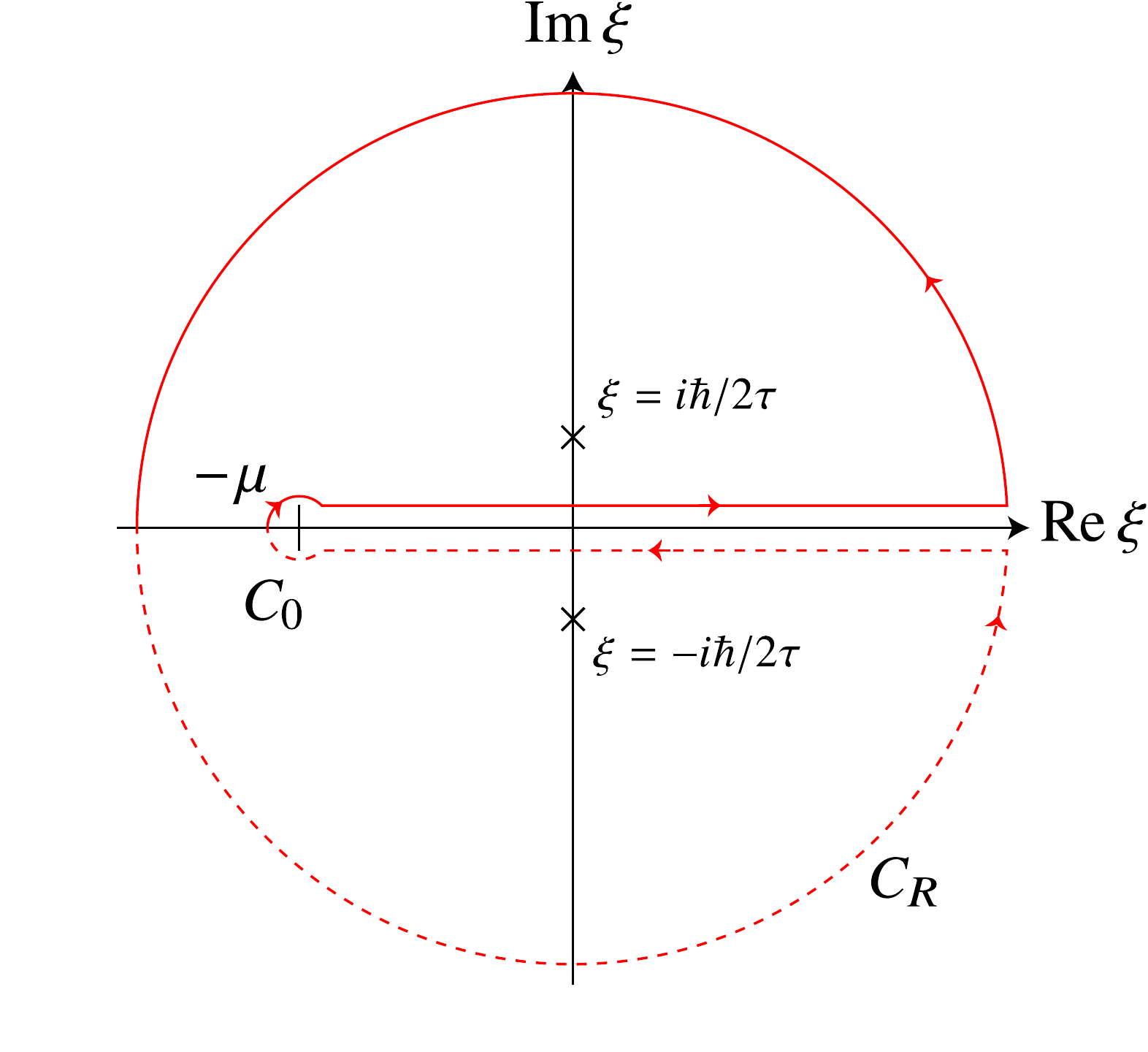}
\caption{\label{fig:exact_path}
The path of the integral is described. %
 The solid and dashed line of the path $C_R$ and $C_0$ denote the pathes on the Riemann surfaces $n = 0$ and $n = 1$, respectively.
The contributions from $C_0$ and $C_R$ vanish in the limit of $R \to \infty$. %
}
\end{figure}

 Considering $\nu (x) \propto x^{1/2}$, the analyticity is as follows:
 Supposed that $z = x + \zi y$ and $w = \sqrt{z} = X + \zi Y$, and in the polar coordinate, $z = r e^{\zi (\theta + 2 n \pi)}$ with $ - \pi \le \theta < \pi$ and $n = 0, \pm 1, \pm 2, \cdots$,
\begin{align}
w
	& = e^{ \frac{1}{2} \log z}
	= e^{\frac{1}{2} \log r + \frac{\zi \theta}{2} + n \pi \zi}
	= \left\{
	\begin{array}{cc}
		\sqrt{r} e^{\zi \frac{\theta}{2}}
	&	(n = 0)
	,\\	- \sqrt{r} e^{\zi \frac{\theta}{2}}
	&	(n = 1)
	,\end{array}
	\right.
\end{align}
where $n = 0, 1$ means the $n$-th Riemann surface: $[-\pi, \pi)$ for $n = 0$, and $[\pi, 3\pi)$ for $n = 1$.
 For the Riemann surface of $n = 0$,
\begin{align}
	X = \sqrt{r} \cos \frac{\theta}{2}
, & \quad
	Y = \sqrt{r} \sin \frac{\theta}{2}
,\end{align}
and, the condition $-\pi \le \theta < \pi$, using $x = r \cos \theta = r (2 \cos^2 (\theta/2) - 1)$, $y = r \sin \theta = 2 r \sin (\theta/2) \cos (\theta/2)$,
\begin{align*}
\cos \frac{\theta}{2}
	= \sqrt{ \frac{1}{2} \left( 1 + \frac{x}{r} \right) }
, \quad
 \sin \frac{\theta}{2}
	= \mathrm{sign}(y) \sqrt{ \frac{1}{2} \left( 1 - \frac{x}{r} \right) }
\end{align*}
 Similarly, for the Riemann surface of $n = 1$,
\begin{align*}
\cos \frac{\theta}{2}
	= -\sqrt{ \frac{1}{2} \left( 1 + \frac{x}{r} \right) }
, \quad
 \sin \frac{\theta}{2}
	= - \mathrm{sign}(y) \sqrt{ \frac{1}{2} \left( 1 - \frac{x}{r} \right) }
\end{align*}
 These are collectively expressed as
\begin{align}
w
	& = \left\{
	\begin{array}{cc}
		\sqrt{x} ( c_{+} (y/x) + \zi \mathrm{sign}(y) c_{-} (y/x) )
	&	(n = 0)
	,\\	- \sqrt{x} ( c_{+} (y/x) + \zi \mathrm{sign}(y) c_{-} (y/x) )
	&	(n = 1)
	,\end{array}
\right.
\label{eq:sqrt_z}
\end{align}
where
\begin{align}
c_{\pm} (\delta)
	& = \sqrt{1 + \delta^2} \left( \frac{\sqrt{1 + \delta^2} \pm 1}{2} \right)^{\tfrac{1}{2}}
\end{align}
 From Eq.~(\ref{eq:sqrt_z}), we can rewrite the path of the integral in Eq.~(\ref{eq:I^A}) as (see Fig.~\ref{fig:exact_path})
\begin{align}
 & \frac{1}{2 \pi} \int_{-\eF}^{\infty} \mathrm{d}\xi \nu (\xi + \eF) \frac{ \nu (\xi + \eF) (\xi + \eF)^l }{(- \xi + \zi \hbar / 2 \tau)^m (- \xi - \zi \hbar / 2 \tau)^n}
\notag \\
	& = \frac{1}{4 \pi} \left(
	\int_{-\eF + \zi 0}^{\infty + \zi 0}
	+ \int_{C_R}
	+ \int_{\infty - \zi 0}^{-\eF - \zi 0}
	+ \int_{C_0}
\right) \mathrm{d}\xi \frac{ \nu (\xi + \eF) (\xi + \eF)^l }{(- \xi + \zi \hbar / 2 \tau)^m (- \xi - \zi \hbar / 2 \tau)^n}
	\notag \\ & = \frac{\zi}{2} \sum_{\eta = \pm 1} \mathrm{Res}_{\xi = \zi \eta \hbar / 2\tau} \left[ \frac{ \nu (\xi + \eF) (\xi + \eF)^l }{(- \xi + \zi \hbar / 2 \tau)^m (- \xi - \zi \hbar / 2 \tau)^n} \right]
\label{eq:integral_general_form}
,\end{align}
where the path $C_R$ is given by $\xi = R e^{\zi \theta}$, $0 \le \theta \le 2 \pi$, (changing the Riemann surfaces at $\theta = \pi$), $C_0$ is given by $\xi = \delta e^{\zi \theta}$, $0 \le \theta < 2 \pi$, and $c_{\pm} = c_{\pm} (\hbar / 2 \eF \tau)$.
 Noted that the sign of the Residue at $\xi = - \zi \hbar / 2 \tau$ is minus, because we gather it of the Riemann surface of $n=1$.

Here, we show the results of the integrals $I_{l m n}$ with $m = 1$.
\begin{subequations}
\begin{align}
I_{0 1 1}
	& = \frac{2 \pi \nu \tau}{\hbar} c_{+}
, \\
I_{0 1 2}
	& = \frac{2 \pi \nu \tau^2}{\hbar^2} \left[
		\left( \zi - \frac{\delta}{2} \frac{ 1 }{1 - \zi \delta} \right) c_{+}
		+ \frac{\delta}{2} \frac{ \zi c_{-} }{1 - \zi \delta}
	\right]
, \\
I_{0 1 3}
	& = - \frac{2 \pi \nu \tau^3}{\hbar^3} \left[
		\left( 1 + \frac{1}{2} \frac{ \zi \delta }{1 - \zi \delta} + \frac{1}{4} \frac{ \delta^2 }{(1 - \zi \delta)^2} \right) c_{+}
		+ \frac{\delta}{2} \frac{ c_{-} }{1 - \zi \delta} \left( 1 - \frac{1}{2} \frac{ \zi \delta }{1 - \zi \delta} \right)
	\right]
, \\
I_{1 1 2}
	& = \frac{2 \pi \nu \eF \tau^2}{\hbar^2} \left[
		\left( \zi - \frac{3 \delta}{2} \right) c_{+}
		+ \frac{\zi \delta}{2} c_{-}
	\right]
, \\
I_{1 1 3}
	& = - \frac{2 \pi \nu \eF \tau^3}{\hbar^3} \left[
		\left(
			1
			+ \frac{3 \zi \delta}{2}
			- \frac{3}{4} \frac{\delta^2}{1 - \zi \delta}
		\right) c_{+}
		+ \frac{\delta}{2} \left(
			1
			+ \frac{3}{2} \frac{\zi \delta}{1 - \zi \delta}
		\right) c_{-}
	\right]
, \\
I_{1 1 4}
	& = \frac{2 \pi \nu \eF \tau^4}{\hbar^4}
		\left[
			\left(
				- \zi
				+ \frac{3 \delta}{2}
				+ \frac{3}{4} \frac{\zi \delta^2}{1 - \zi \delta}
				+ \frac{1}{4} \frac{\delta^3}{( 1 - \zi \delta)^2}
			\right) c_{+}
			- \frac{\delta}{2} \left(
				\zi
				- \frac{3}{2} \frac{\delta}{1 - \zi \delta}
				+ \frac{1}{2} \frac{\zi \delta^2}{( 1 - \zi \delta)^2}
				\right) c_{-}
		\right]
, \\
I_{2 1 4}
	& = \frac{2 \pi \nu \eF^2 \tau^4}{\hbar^4} \left[
		\left(
			- \zi
			+ \frac{5 \delta}{2}
			+ \frac{9 \zi \delta^2}{4}
			- \frac{5}{4} \frac{\delta^3}{1 - \zi \delta}
		\right) c_{+}
		- \frac{\delta}{2} \left(
			\zi 
			 - \frac{5 \delta}{2}
			 - \frac{5 \zi \delta^2}{2} \frac{1}{1 - \zi \delta}
		\right) c_{-}
	\right]
, \\
I_{2 1 5}
	& = \frac{2 \pi \nu \eF^2 \tau^5}{\hbar^5} \left[
		\left(
			1
			+ \frac{5 \zi \delta}{2}
			- \frac{9 \delta^2}{4}
			- \frac{5}{4} \frac{\zi \delta^3}{1 - \zi \delta}
			- \frac{5}{16} \frac{\delta^4}{(1 - \zi \delta)^2}
		\right) c_{+}
		+ \frac{\delta}{2} \left(
			1
			+ \frac{5 \zi \delta}{2}
			- \frac{5}{2} \frac{\delta^2}{1 - \zi \delta}
			+ \frac{5}{8} \frac{\zi \delta^3}{(1 - \zi \delta)^2}
		\right) c_{-}
	\right]
,\end{align}
\label{eqs:I_lmn_evaluated}%
\end{subequations}
where $\delta = \hbar / 2 \eF \tau \, (\ll 1)$.
For small $\delta$ $(> 0)$,
\begin{subequations}
\begin{align}
c_{+} (\delta)
& = 1 + (5/8) \delta^2 - (13/128) \delta^4 + \cdots
, \\
c_{-} (\delta)
& = \delta/2 + (3/16) \delta^3 - (17/256) \delta^5 - \cdots
\end{align}

\label{eq:small_x}
\end{subequations}

%
\bibliography{reference,/home/gt/References/15,/home/gt/References/gt,/home/gt/References/remarks}

\begin{thebibliography}{23}%
\makeatletter
\providecommand \@ifxundefined [1]{%
 \@ifx{#1\undefined}
}%
\providecommand \@ifnum [1]{%
 \ifnum #1\expandafter \@firstoftwo
 \else \expandafter \@secondoftwo
 \fi
}%
\providecommand \@ifx [1]{%
 \ifx #1\expandafter \@firstoftwo
 \else \expandafter \@secondoftwo
 \fi
}%
\providecommand \natexlab [1]{#1}%
\providecommand \enquote  [1]{``#1''}%
\providecommand \bibnamefont  [1]{#1}%
\providecommand \bibfnamefont [1]{#1}%
\providecommand \citenamefont [1]{#1}%
\providecommand \href@noop [0]{\@secondoftwo}%
\providecommand \href [0]{\begingroup \@sanitize@url \@href}%
\providecommand \@href[1]{\@@startlink{#1}\@@href}%
\providecommand \@@href[1]{\endgroup#1\@@endlink}%
\providecommand \@sanitize@url [0]{\catcode `\\12\catcode `\$12\catcode
  `\&12\catcode `\#12\catcode `\^12\catcode `\_12\catcode `\%12\relax}%
\providecommand \@@startlink[1]{}%
\providecommand \@@endlink[0]{}%
\providecommand \url  [0]{\begingroup\@sanitize@url \@url }%
\providecommand \@url [1]{\endgroup\@href {#1}{\urlprefix }}%
\providecommand \urlprefix  [0]{URL }%
\providecommand \Eprint [0]{\href }%
\providecommand \doibase [0]{http://dx.doi.org/}%
\providecommand \selectlanguage [0]{\@gobble}%
\providecommand \bibinfo  [0]{\@secondoftwo}%
\providecommand \bibfield  [0]{\@secondoftwo}%
\providecommand \translation [1]{[#1]}%
\providecommand \BibitemOpen [0]{}%
\providecommand \bibitemStop [0]{}%
\providecommand \bibitemNoStop [0]{.\EOS\space}%
\providecommand \EOS [0]{\spacefactor3000\relax}%
\providecommand \BibitemShut  [1]{\csname bibitem#1\endcsname}%
\let\auto@bib@innerbib\@empty
\bibitem [{\citenamefont {Silsbee}\ \emph {et~al.}(1979)\citenamefont
  {Silsbee}, \citenamefont {Janossy},\ and\ \citenamefont
  {Monod}}]{Silsbee1979}%
  \BibitemOpen
  \bibfield  {author} {\bibinfo {author} {\bibfnamefont {R.~H.}\ \bibnamefont
  {Silsbee}}, \bibinfo {author} {\bibfnamefont {A.}~\bibnamefont {Janossy}}, \
  and\ \bibinfo {author} {\bibfnamefont {P.}~\bibnamefont {Monod}},\ }\href
  {\doibase 10.1103/PhysRevB.19.4382} {\bibfield  {journal} {\bibinfo
  {journal} {Phys. Rev. B}\ }\textbf {\bibinfo {volume} {19}},\ \bibinfo
  {pages} {4382} (\bibinfo {year} {1979})}\BibitemShut {NoStop}%
\bibitem [{\citenamefont {Mizukami}\ \emph {et~al.}(2001)\citenamefont
  {Mizukami}, \citenamefont {Ando},\ and\ \citenamefont
  {Miyazaki}}]{Mizukami2001}%
  \BibitemOpen
  \bibfield  {author} {\bibinfo {author} {\bibfnamefont {S.}~\bibnamefont
  {Mizukami}}, \bibinfo {author} {\bibfnamefont {Y.}~\bibnamefont {Ando}}, \
  and\ \bibinfo {author} {\bibfnamefont {T.}~\bibnamefont {Miyazaki}},\ }\href
  {\doibase 10.1016/S0304-8853(00)01097-0} {\bibfield  {journal} {\bibinfo
  {journal} {J. Magn. Magn. Mater.}\ }\textbf {\bibinfo {volume} {226-230}},\
  \bibinfo {pages} {1640} (\bibinfo {year} {2001})}\BibitemShut {NoStop}%
\bibitem [{\citenamefont {Tserkovnyak}\ \emph {et~al.}(2002)\citenamefont
  {Tserkovnyak}, \citenamefont {Brataas},\ and\ \citenamefont
  {Bauer}}]{Tserkovnyak2002}%
  \BibitemOpen
  \bibfield  {author} {\bibinfo {author} {\bibfnamefont {Y.}~\bibnamefont
  {Tserkovnyak}}, \bibinfo {author} {\bibfnamefont {A.}~\bibnamefont
  {Brataas}}, \ and\ \bibinfo {author} {\bibfnamefont {G.~E.~W.}\ \bibnamefont
  {Bauer}},\ }\href {\doibase 10.1103/PhysRevLett.88.117601} {\bibfield
  {journal} {\bibinfo  {journal} {Phys. Rev. Lett.}\ }\textbf {\bibinfo
  {volume} {88}},\ \bibinfo {pages} {117601} (\bibinfo {year}
  {2002})}\BibitemShut {NoStop}%
\bibitem [{\citenamefont {Costache}\ \emph {et~al.}(2006)\citenamefont
  {Costache}, \citenamefont {Sladkov}, \citenamefont {Watts}, \citenamefont
  {{van der Wal}},\ and\ \citenamefont {{van Wees}}}]{Costache2006}%
  \BibitemOpen
  \bibfield  {author} {\bibinfo {author} {\bibfnamefont {M.~V.}\ \bibnamefont
  {Costache}}, \bibinfo {author} {\bibfnamefont {M.}~\bibnamefont {Sladkov}},
  \bibinfo {author} {\bibfnamefont {S.~M.}\ \bibnamefont {Watts}}, \bibinfo
  {author} {\bibfnamefont {C.~H.}\ \bibnamefont {{van der Wal}}}, \ and\
  \bibinfo {author} {\bibfnamefont {B.~J.}\ \bibnamefont {{van Wees}}},\ }\href
  {\doibase 10.1103/PhysRevLett.97.216603} {\bibfield  {journal} {\bibinfo
  {journal} {Phys. Rev. Lett.}\ }\textbf {\bibinfo {volume} {97}},\ \bibinfo
  {pages} {216603} (\bibinfo {year} {2006})}\BibitemShut {NoStop}%
\bibitem [{\citenamefont {Saitoh}\ \emph {et~al.}(2006)\citenamefont {Saitoh},
  \citenamefont {Ueda}, \citenamefont {Miyajima},\ and\ \citenamefont
  {Tatara}}]{Saitoh2006}%
  \BibitemOpen
  \bibfield  {author} {\bibinfo {author} {\bibfnamefont {E.}~\bibnamefont
  {Saitoh}}, \bibinfo {author} {\bibfnamefont {M.}~\bibnamefont {Ueda}},
  \bibinfo {author} {\bibfnamefont {H.}~\bibnamefont {Miyajima}}, \ and\
  \bibinfo {author} {\bibfnamefont {G.}~\bibnamefont {Tatara}},\ }\href
  {\doibase 10.1063/1.2199473} {\bibfield  {journal} {\bibinfo  {journal}
  {Appl. Phys. Lett.}\ }\textbf {\bibinfo {volume} {88}},\ \bibinfo {pages}
  {182509} (\bibinfo {year} {2006})}\BibitemShut {NoStop}%
\bibitem [{\citenamefont {Edelstein}(1990)}]{Edelstein90}%
  \BibitemOpen
  \bibfield  {author} {\bibinfo {author} {\bibfnamefont {V.}~\bibnamefont
  {Edelstein}},\ }\href {\doibase 10.1016/0038-1098(90)90963-C} {\bibfield
  {journal} {\bibinfo  {journal} {Solid State Communications}\ }\textbf
  {\bibinfo {volume} {73}},\ \bibinfo {pages} {233 } (\bibinfo {year}
  {1990})}\BibitemShut {NoStop}%
\bibitem [{\citenamefont {S\'anchez}\ \emph {et~al.}(2013)\citenamefont
  {S\'anchez}, \citenamefont {Vila}, \citenamefont {Desfonds}, \citenamefont
  {Gambarelli}, \citenamefont {Attan\'e}, \citenamefont {De~Teresa},
  \citenamefont {Mag\'en},\ and\ \citenamefont {Fert}}]{Sanchez2013}%
  \BibitemOpen
  \bibfield  {author} {\bibinfo {author} {\bibfnamefont {J.~C.~R.}\
  \bibnamefont {S\'anchez}}, \bibinfo {author} {\bibfnamefont {L.}~\bibnamefont
  {Vila}}, \bibinfo {author} {\bibfnamefont {G.}~\bibnamefont {Desfonds}},
  \bibinfo {author} {\bibfnamefont {S.}~\bibnamefont {Gambarelli}}, \bibinfo
  {author} {\bibfnamefont {J.~P.}\ \bibnamefont {Attan\'e}}, \bibinfo {author}
  {\bibfnamefont {J.~M.}\ \bibnamefont {De~Teresa}}, \bibinfo {author}
  {\bibfnamefont {C.}~\bibnamefont {Mag\'en}}, \ and\ \bibinfo {author}
  {\bibfnamefont {A.}~\bibnamefont {Fert}},\ }\href {\doibase
  10.1038/ncomms3944} {\bibfield  {journal} {\bibinfo  {journal} {Nat.
  Commun.}\ }\textbf {\bibinfo {volume} {4}},\ \bibinfo {pages} {2944}
  (\bibinfo {year} {2013})}\BibitemShut {NoStop}%
\bibitem [{\citenamefont {Volovik}(1987)}]{Volovik1987}%
  \BibitemOpen
  \bibfield  {author} {\bibinfo {author} {\bibfnamefont {G.~E.}\ \bibnamefont
  {Volovik}},\ }\href {\doibase 10.1088/0022-3719/20/7/003} {\bibfield
  {journal} {\bibinfo  {journal} {J. Phys. C Solid State Phys.}\ }\textbf
  {\bibinfo {volume} {20}},\ \bibinfo {pages} {L83} (\bibinfo {year} {3月
  1987})}\BibitemShut {NoStop}%
\bibitem [{\citenamefont {Tatara}(2019)}]{TataraReview19}%
  \BibitemOpen
  \bibfield  {author} {\bibinfo {author} {\bibfnamefont {G.}~\bibnamefont
  {Tatara}},\ }\href {\doibase https://doi.org/10.1016/j.physe.2018.05.011}
  {\bibfield  {journal} {\bibinfo  {journal} {Physica E: Low-dimensional
  Systems and Nanostructures}\ }\textbf {\bibinfo {volume} {106}},\ \bibinfo
  {pages} {208 } (\bibinfo {year} {2019})}\BibitemShut {NoStop}%
\bibitem [{\citenamefont {Kim}\ \emph {et~al.}(2012)\citenamefont {Kim},
  \citenamefont {Moon}, \citenamefont {Lee},\ and\ \citenamefont
  {Lee}}]{Kim2012}%
  \BibitemOpen
  \bibfield  {author} {\bibinfo {author} {\bibfnamefont {K.-W.}\ \bibnamefont
  {Kim}}, \bibinfo {author} {\bibfnamefont {J.-H.}\ \bibnamefont {Moon}},
  \bibinfo {author} {\bibfnamefont {K.-J.}\ \bibnamefont {Lee}}, \ and\
  \bibinfo {author} {\bibfnamefont {H.-W.}\ \bibnamefont {Lee}},\ }\href
  {\doibase 10.1103/PhysRevLett.108.217202} {\bibfield  {journal} {\bibinfo
  {journal} {Phys. Rev. Lett.}\ }\textbf {\bibinfo {volume} {108}},\ \bibinfo
  {pages} {217202} (\bibinfo {year} {2012})}\BibitemShut {NoStop}%
\bibitem [{\citenamefont {Takeuchi}\ and\ \citenamefont
  {Tatara}(2012)}]{Takeuchi2012}%
  \BibitemOpen
  \bibfield  {author} {\bibinfo {author} {\bibfnamefont {A.}~\bibnamefont
  {Takeuchi}}\ and\ \bibinfo {author} {\bibfnamefont {G.}~\bibnamefont
  {Tatara}},\ }\href {\doibase 10.1143/JPSJ.81.033705} {\bibfield  {journal}
  {\bibinfo  {journal} {J. Phys. Soc. Jpn.}\ }\textbf {\bibinfo {volume}
  {81}},\ \bibinfo {pages} {033705} (\bibinfo {year} {2月 2012})}\BibitemShut
  {NoStop}%
\bibitem [{\citenamefont {Nakabayashi}\ and\ \citenamefont
  {Tatara}(2014)}]{Nakabayashi2014}%
  \BibitemOpen
  \bibfield  {author} {\bibinfo {author} {\bibfnamefont {N.}~\bibnamefont
  {Nakabayashi}}\ and\ \bibinfo {author} {\bibfnamefont {G.}~\bibnamefont
  {Tatara}},\ }\href {\doibase 10.1088/1367-2630/16/1/015016} {\bibfield
  {journal} {\bibinfo  {journal} {New J. Phys.}\ }\textbf {\bibinfo {volume}
  {16}},\ \bibinfo {pages} {015016} (\bibinfo {year} {2014})}\BibitemShut
  {NoStop}%
\bibitem [{\citenamefont {Tatara}\ \emph {et~al.}(2013)\citenamefont {Tatara},
  \citenamefont {Nakabayashi},\ and\ \citenamefont {Lee}}]{Tatara_smf13}%
  \BibitemOpen
  \bibfield  {author} {\bibinfo {author} {\bibfnamefont {G.}~\bibnamefont
  {Tatara}}, \bibinfo {author} {\bibfnamefont {N.}~\bibnamefont {Nakabayashi}},
  \ and\ \bibinfo {author} {\bibfnamefont {K.-J.}\ \bibnamefont {Lee}},\ }\href
  {\doibase 10.1103/PhysRevB.87.054403} {\bibfield  {journal} {\bibinfo
  {journal} {Phys. Rev. B}\ }\textbf {\bibinfo {volume} {87}},\ \bibinfo
  {pages} {054403} (\bibinfo {year} {2013})}\BibitemShut {NoStop}%
\bibitem [{\citenamefont {Takeuchi}\ \emph {et~al.}(2010)\citenamefont
  {Takeuchi}, \citenamefont {Hosono},\ and\ \citenamefont
  {Tatara}}]{Takeuchi10}%
  \BibitemOpen
  \bibfield  {author} {\bibinfo {author} {\bibfnamefont {A.}~\bibnamefont
  {Takeuchi}}, \bibinfo {author} {\bibfnamefont {K.}~\bibnamefont {Hosono}}, \
  and\ \bibinfo {author} {\bibfnamefont {G.}~\bibnamefont {Tatara}},\ }\href
  {\doibase 10.1103/PhysRevB.81.144405} {\bibfield  {journal} {\bibinfo
  {journal} {Phys. Rev. B}\ }\textbf {\bibinfo {volume} {81}},\ \bibinfo
  {pages} {144405} (\bibinfo {year} {2010})}\BibitemShut {NoStop}%
\bibitem [{\citenamefont {Nakazawa}\ and\ \citenamefont
  {Kohno}(2014)}]{Nakazawa2014}%
  \BibitemOpen
  \bibfield  {author} {\bibinfo {author} {\bibfnamefont {K.}~\bibnamefont
  {Nakazawa}}\ and\ \bibinfo {author} {\bibfnamefont {H.}~\bibnamefont
  {Kohno}},\ }\href {\doibase 10.7566/JPSJ.83.073707} {\bibfield  {journal}
  {\bibinfo  {journal} {J. Phys. Soc. Jpn.}\ }\textbf {\bibinfo {volume}
  {83}},\ \bibinfo {pages} {073707} (\bibinfo {year} {2014})}\BibitemShut
  {NoStop}%
\bibitem [{\citenamefont {Tatara}(2018)}]{Tatara18}%
  \BibitemOpen
  \bibfield  {author} {\bibinfo {author} {\bibfnamefont {G.}~\bibnamefont
  {Tatara}},\ }\href {\doibase 10.1103/PhysRevB.98.174422} {\bibfield
  {journal} {\bibinfo  {journal} {Phys. Rev. B}\ }\textbf {\bibinfo {volume}
  {98}},\ \bibinfo {pages} {174422} (\bibinfo {year} {2018})}\BibitemShut
  {NoStop}%
\bibitem [{\citenamefont {Kubo}\ \emph {et~al.}(1991)\citenamefont {Kubo},
  \citenamefont {Toda},\ and\ \citenamefont {Hashitsume}}]{Kubo1991}%
  \BibitemOpen
  \bibfield  {author} {\bibinfo {author} {\bibfnamefont {P.~D.~R.}\
  \bibnamefont {Kubo}}, \bibinfo {author} {\bibfnamefont {P.~D.~M.}\
  \bibnamefont {Toda}}, \ and\ \bibinfo {author} {\bibfnamefont {P.~D.~N.}\
  \bibnamefont {Hashitsume}},\ }in\ \href {\doibase
  10.1007/978-3-642-58244-8_4} {\emph {\bibinfo {booktitle} {Statistical
  {{Physics II}}}}},\ \bibinfo {series and number} {\bibinfo {series} {Springer
  Series in Solid-State Sciences}\ No.~\bibinfo {number} {31}}\ (\bibinfo
  {publisher} {{Springer Berlin Heidelberg}},\ \bibinfo {year} {1991})\ pp.\
  \bibinfo {pages} {146--202}\BibitemShut {NoStop}%
\bibitem [{\citenamefont {Jujo}(2006)}]{Jujo2006}%
  \BibitemOpen
  \bibfield  {author} {\bibinfo {author} {\bibfnamefont {T.}~\bibnamefont
  {Jujo}},\ }\href {\doibase 10.1143/JPSJ.75.104709} {\bibfield  {journal}
  {\bibinfo  {journal} {J. Phys. Soc. Jpn.}\ }\textbf {\bibinfo {volume}
  {75}},\ \bibinfo {pages} {104709} (\bibinfo {year} {2006})}\BibitemShut
  {NoStop}%
\bibitem [{\citenamefont {Kohno}\ and\ \citenamefont
  {Shibata}(2007)}]{Kohno2007}%
  \BibitemOpen
  \bibfield  {author} {\bibinfo {author} {\bibfnamefont {H.}~\bibnamefont
  {Kohno}}\ and\ \bibinfo {author} {\bibfnamefont {J.}~\bibnamefont
  {Shibata}},\ }\href {\doibase 10.1143/JPSJ.76.063710} {\bibfield  {journal}
  {\bibinfo  {journal} {J. Phys. Soc. Jpn.}\ }\textbf {\bibinfo {volume}
  {76}},\ \bibinfo {pages} {063710} (\bibinfo {year} {2007})}\BibitemShut
  {NoStop}%
\bibitem [{\citenamefont {Zagoskin}(1998)}]{Zagoskin1998}%
  \BibitemOpen
  \bibfield  {author} {\bibinfo {author} {\bibfnamefont {A.~M.}\ \bibnamefont
  {Zagoskin}},\ }\href@noop {} {\emph {\bibinfo {title} {Quantum {{Theory}} of
  {{Many}}-{{Body Systems}}: {{Techniques}} and {{Applications}}}}}\ (\bibinfo
  {publisher} {{Springer- Verlag}},\ \bibinfo {year} {1998})\BibitemShut
  {NoStop}%
\bibitem [{\citenamefont {Rickayzen}(2013)}]{Rickayzen2013}%
  \BibitemOpen
  \bibfield  {author} {\bibinfo {author} {\bibfnamefont {G.}~\bibnamefont
  {Rickayzen}},\ }\href@noop {} {\emph {\bibinfo {title} {Green's {{Functions}}
  and {{Condensed Matter}}}}},\ \bibinfo {edition} {reprint}\ ed.\ (\bibinfo
  {publisher} {{Dover Publications}},\ \bibinfo {address} {New York},\ \bibinfo
  {year} {2013})\BibitemShut {NoStop}%
\bibitem [{Note1()}]{Note1}%
  \BibitemOpen
  \bibinfo {note} {As we have introduced the convergence factor $\eta $ as in
  Eq.~(\ref {eq:convergence_factor}), $\varDelta _1 B (t)$ is also assumed to
  be expressed as $\varDelta _1 B (t) = e^{\eta t} \DOTSI \intop \ilimits@
  \varDelta _1 B (\omega ) e^{- i\omega t} \protect \mathrm {d}\omega /2 \pi $
  for the time-translational symmetry.}\BibitemShut {Stop}%
\bibitem [{Note2()}]{Note2}%
  \BibitemOpen
  \bibinfo {note} {Considering the convergence factors, we have assumed
  $\varDelta _2 B (t) = e^{(\eta + \eta ') t} \DOTSI \intop \ilimits@ \varDelta
  _2 B (\omega ) e^{-i\omega t} \protect \mathrm {d}t/2 \pi $}\BibitemShut
  {NoStop}%
\end{thebibliography}%
\end{document}